\documentclass{jaa}
\usepackage{natbib}
\usepackage{newtxtext,newtxmath}
\usepackage{graphicx}
\usepackage{amsmath}
\usepackage{caption}
\usepackage{subcaption}
\usepackage{float}
\usepackage[export]{adjustbox}
\usepackage{natbib}

\usepackage{amsmath} 

\def\ao{Applied\ Optics}

\def\aap{Astron.\ Astrophys.}
\def\apjl{Astrophys.\ J.\ Lett.}

\def\apjs{ApJS}

\usepackage{graphicx}
\usepackage[colorlinks=true, allcolors=blue]{hyperref}

\usepackage{soul,xcolor}
\setstcolor{magenta}

\begin{document}\sloppy

\title{Turbulence and Magnetic Fields in Star Formation}



\author{Archana Soam$^{1,*}$,
Chakali Eswaraiah$^{2}$,
Amit Seta$^{3}$,
Lokesh Dewangan$^{4}$,
and Maheswar G.$^{1}$}
\affilOne{\textsuperscript{1} Indian Institute of Astrophysics, II Block, Koramangala, Bengaluru 560034, India\\}
\affilTwo{\textsuperscript{2} Indian Institute of Science Education and Research (IISER) Tirupati, Rami Reddy Nagar, Karakambadi Road, Mangalam (P.O.), 517507, Tirupati, India\\}
\affilThree{\textsuperscript{3} Research School of Astronomy and Astrophysics, Australian National University, Canberra, ACT 2611, Australia\\}
\affilFour{\textsuperscript{4}Astronomy and Astrophysics Division, Physical Research Laboratory, 380009, Navrangpura, Ahmedabad, India\\}


\twocolumn[{

\maketitle

\corres{archana.soam@iiap.res.in}


\begin{abstract}
Molecular clouds are prime locations to study the process of star formation. These clouds contain filamentary structures and cores, which are crucial sites for the formation of young stars. The star-formation process has been investigated using various techniques, including polarimetry for tracing magnetic fields. In this small review-cum-short report, we put together the efforts (mainly from the Indian community) to understand the roles of turbulence and magnetic fields in star formation. These are two components of the ISM competing against gravity, which is primarily responsible for the collapse of gas to form stars. We also include attempts made using simulations of molecular clouds to study this competition. Studies on feedback and magnetic fields are combined and listed to understand the importance of the interaction between two energies in setting the current observed star formation efficiency. We have listed available and upcoming facilities with the polarization capabilities needed to trace magnetic fields. We have also stated the importance of ongoing and desired collaborations between Indian communities and facilities abroad to shed more light on the roles of turbulence and magnetic fields in the process of star formation.
---------
\end{abstract}


}]


\doinum{}
\artcitid{\#\#\#\#}
\volnum{000}
\year{0000}
\pgrange{1--}
\setcounter{page}{1}
\lp{1}

\section{Introduction, findings, and global efforts}

Interstellar Medium (ISM) is a dynamic medium between the stars which consists of thermal gas, dust, cosmic rays, turbulence, and magnetic fields. Stars form in cold, dense, and gravitationally unstable regions, a.k.a., molecular clouds (e.g., \citealt{benson&Myers1989}). Star formation is not the end-point process of ISM evolution; feedback effects from young stars and supernovae play a significant role in molecular cloud chemistry, evolution, and driving ISM turbulence (e.g., \citealt{krumholz2014}). 

\begin{figure*} 
\centering
\resizebox{8cm}{8cm}{\includegraphics{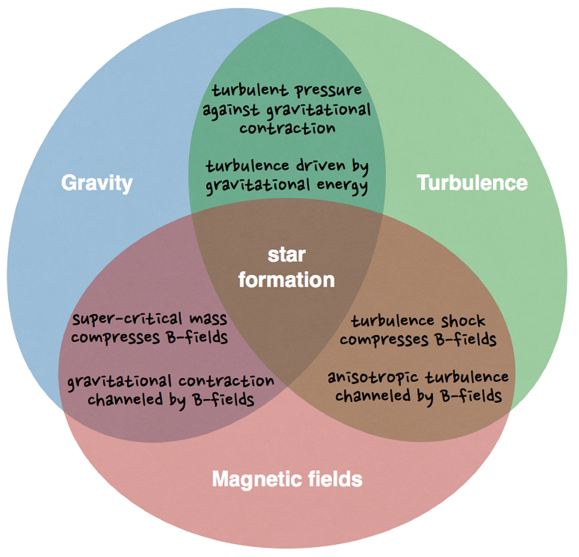}}
\caption{Venn diagram (adopted from \citealt{2017tcsf.book.....L}) depicting the complex interplay between B-fields, gravity, and turbulence, which are the three major players of star formation.}\label{fig:energies}
\end{figure*}

Turbulence and Magnetic fields (B-fields) are found to play important roles in the formation and evolution of these molecular clouds (see \citealt{pattle2022}). B-field pervades the entire interstellar medium composed of different phases in the Galaxy \citep{fer2020} such as warm neutral medium (WNM), cold neutral medium (CNM), cold molecular medium (CMM), and hot ionized medium (HIM). Both recent simulations \citep{SF2022, GentEA2023} and observations \citep{BraccoEA2020, BorlaffEA2021} show that the properties of B-fields differ with the phase of the ISM. In the diffuse interstellar medium (cold neutral medium; CNM), the magnetic pressure ($\approx$10,000 K cm$^{-3}$; with B-field strength of $\approx$ 6 $\mu$G) contributes as much as the turbulence does (see \citealt{fer2020} for temperature, density, and velocity information in different phases of the ISM) but dominates over thermal pressure ($\approx$4000 K cm$^{-3}$; \citealt{JenkinsTripp2011}). How does the contribution of each pressure component change from low- to high-density (or large- to small-scale) regions of molecular clouds, and the relative importance of each component against another is a matter of investigation. B-field couples with the gas and dust \citep{mestel1966}, and governs various physical processes such as (i) cloud formation and evolution, (ii) fragmentation and collapse of filaments into dense cores and protostars, and (iii) formation of outflows and circumstellar disks. B-field, rather than playing a sole role, interacts with other key agents such as gravity and turbulence in a complex fashion as shown in Figure \ref{fig:energies} and dictates the star formation process.

It is interesting to investigate the roles of turbulence and B-fields in the evolution of ISM. For that, we can use the virial theorem to discuss the relative importance of gravitational, kinematic, and magnetic energies in any region. The Venn diagram shown in Figure \ref{fig:energies} represents the importance of these energies. Even after a lot of understanding in this direction, there are still debates on certain issues regarding the importance of turbulence and magnetic fields. Along with this, some questions must be addressed for a complete understanding of the ISM evolution. These questions are:
\begin{itemize}
    \item why the observed star formation is so inefficient when compared to theories and models?
    \item how do turbulence and B-fields help in regulating gas motions, and how strong/weak is this interaction as we go to different density or spatial regimes?
    \item how do the morphology of B-fields change on different spatial scales, and how does their strength vary on these scales (i.e.~from parsec to sub-parsec scales)?
    \item how does the energy budget of star-forming regions look like? How do the gravitational, kinematic, and magnetic energies of the starless/star-forming cores compete with each other?
\end{itemize}

Turbulence plays a dual role in the formation of stars. While it makes it harder for gravity to form stars by moving the gas around, the turbulence in the colder, denser ISM is supersonic and thus creates strong local gas compressions due to shocks, which can act as potential sites of star formation. Thus, the turbulence acts as a `necessary evil'. 
Turbulence decays from large to small scales in the ISM. \citet{larson1981} found an empirical, positive correlation between the physical size scale and the
velocity dispersion of molecular clouds where stars form. Although this
the study was reported based on measurements from a heterogeneous sample of clouds, the correlation was taken as an indication that turbulence dissipates towards smaller spatial scales where gravitational collapse happens and star formation takes place \citep{maclow2004, krumholz2005, elmegreen2004, hennebelle2008} and references therein. The formed stars, via stellar feedback (due to stellar winds and supernova explosions from the deaths of massive stars), eject material back into the ISM and replenish turbulence.  The turbulence can also be driven by other mechanisms such as accretion, spiral shocks, galactic shear, magnetorotational instability, and more generally gravity-driven turbulence \citep{KEA2018}. 

The star formation process also depends on the nature of turbulence, as has been shown via numerical simulations \citep{FK2012}, that the compressive modes (curl-free, expected for shock-driven turbulence) of turbulence give a higher (by order of magnitude) star formation rate in comparison to solenoidal modes (divergence-free, expected for turbulence driven by galactic shear and magnetorotational instability). This is due to compressive modes creating regions of stronger densities, which makes it easier to initiate collapse, eventually leading to the formation of stars.

In a study by \citet{Orkisz2017}, turbulence and star formation efficiency in Orion B, are studied observationally. The nature of the turbulence controls star formation efficiency i.e., compressive motions, as opposed to solenoidal motions, can trigger the collapse of cores, or mark the expansion of HII regions. This work derived the fractions of momentum density contained in the solenoidal and compressive modes of turbulence in the Orion B molecular cloud. They found in Orion B, there can be a strong intra-cloud variability of the compressive and solenoidal fractions, these fractions being in turn related to the star formation efficiency.
\citet{Hayward2017} published work on stellar feedback regulating star formation. An analytic model was presented on momentum deposition from stellar feedback and simultaneous regulation of star formation in a turbulent interstellar medium. High-density patches in the ISM can be ‘pushed’ by feedback, thereby driving turbulence and self-regulating local star formation. We will expand more below on stellar feedback and star formation.

The star formation process is highly inefficient, i.e., only a small fraction ($\lesssim 10\%$) of the gas in molecular clouds is converted to stars, and the star formation rate (SFR) is of the order of one solar mass per year \citep{KT2007}. The high-energy radiation and stellar winds inject substantial energy into the ISM, which affects the dynamic and chemical evolution of the ISM. The infrared structures associated with HII regions are promising sites to investigate the feedback processes of massive OB stars \citep{anderson2011, anderson2014}. This question of inefficient star formation might be addressed by investigation the possible suppression due to the feedback process (e.g., \citealt{MacLow2017}, \citealt{semadeni2019}). However, simulations find this very complicated to handle, and hence it is difficult to understand the role of feedback in regulating star formation (e.g., \citealt{dale2014}, \citealt{geen2020}, \citealt{kim2021}).

Many recent studies published with the help of ALMA and SMA polarization measurements have shed a lot of light on the regulation of star formation by B-fields. In a study by \citet{Dall'Olio2019}, they determined the
B-field morphology and strength in the high-mass starforming region G9.62+0.19 to investigate its relation in the evolutionary
sequence of the cores. They found high magnetic field strength and smooth polarised emission which highlights the importance of B-fields in fragmentation and the collapse process in G9.62+0.19 and magnetised core evolution. Many such studies are published using ALMA (e.g.,~\citealt{Girart2018, Alves2018}). Polarization observations from SMA for a Legacy Survey of 18 massive dense clumps were performed by \citet{Zhang2014}. They found a strong correlation between the fragmentation level with the density of the parental clump and a trend of fragmentation with the mass-to-flux ratio. \citet{Beltran2019} used ALMA observations to reveal the hour-glass morphology of magnetic fields in a hot molecular core G31.41+0.31 on a scale of $<$1000 AU. Thus, both the strength and structure of the ISM magnetic field play a role in the formation of stars. B-fields also play an important role in the structural evolution of molecular clouds, such as the ones found on the periphery of expanding HII regions. The details are given in the following paragraph.

\begin{figure*} 
\centering
\resizebox{18cm}{9cm}{\includegraphics{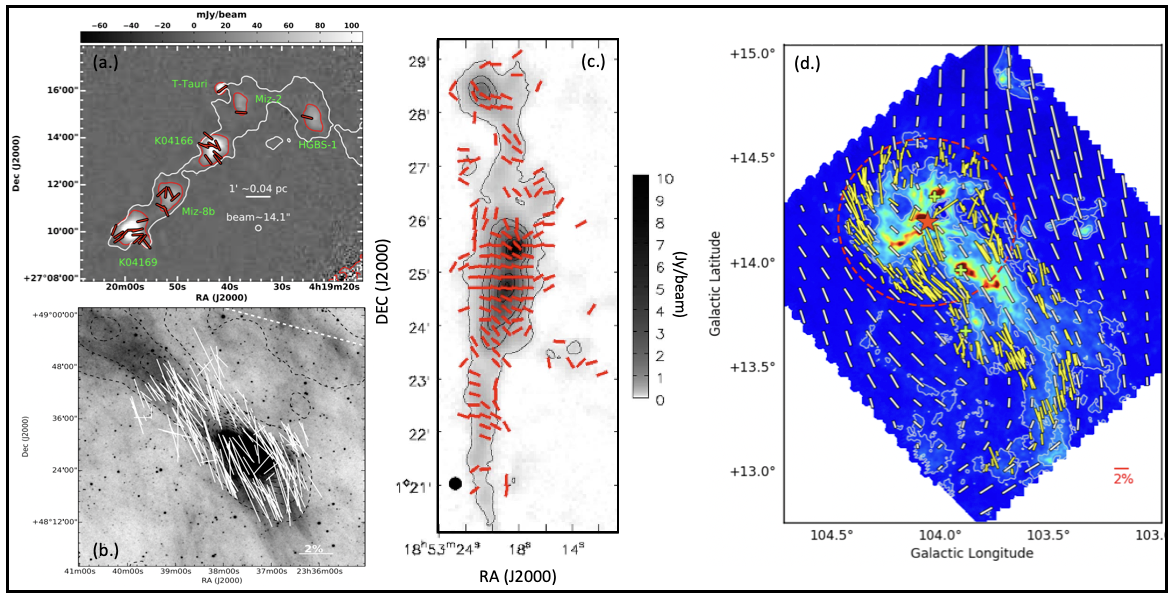}}
\caption{A compilations of B-field morphologies traced in different regions of the ISM. These regions include (a.) Taurus cores \citep{eswar2021}, (b.) Cometary Globule (CG) Gal 110-13 \citep{neha2016}, (c.) filament G34.43+0.24 \citep{soam2019}, and (d.) bright-rimmed cloud (BRC) L1172/1174 cloud complex \citep{saha2021}.}\label{fig:MF_all}
\end{figure*}

The photoionized gas and/or stellar winds associated with the OB stars are also often considered as the major contributor for the origin of the infrared bright dusty structures. Another possible mechanism responsible for the infrared bright structures could be supernova explosions. On the other hand, the existence of a ring-like morphology without HII regions is unlikely to be explained by the feedback of OB stars. Magnetic field morphologies in such structures have also been studied (see panels b. and d. of Figure \ref{fig:MF_all}). Recently, \citep{pavel2012} examined the interaction of HII region driven Galactic bubbles with the Galactic magnetic field and found that external magnetic fields are important during the earliest phases of the evolution of HII region driven Galactic bubbles. \citet{eswar2017}, based on NIR polarimetry, have showed that B-fields not only guide the ionization fronts from the embedded HII region RCW 57A thereby aiding the large scale bipolar bubbles. In addition \citet{eswar2020} have traced compressed B-fields around two massive clumps of Sh2-201. The B-fields configured into bow-like morphology are found to strong enough to shield the clumps from erosion by HII regio and stabilize the clumps against gravitational collapse.. \citet{tahani2023} study the sub-mm polarization of HII regions  (bubbles) associated with the NGC 6334 molecular cloud. They found that the gas and magnetic field lines have been pushed away from the bubble, toward an almost tangential (to the bubble) magnetic field morphology. They also noticed a radial polarization pattern in one of the bubbles. However, to our knowledge, a detailed observational study of the influence of the magnetic field on bubbles is still very limited in the literature.


 
Despite the primary importance of B-fields in the star formation process, the relative importance of magnetic fields in comparison to other factors is not fully understood.  This is partly owing to the lack of direct B-field strength measurements (based on the Zeeman effect; \citealt{Crutcher&Richard2019}) and the techniques used for measuring the magnetic fields. Therefore, it has always been relatively easier to constrain the gravity and turbulence from the column density and molecular lines data, respectively. With the advent of more sensitive and wide-field polarimeters, it is now possible to trace the magnetic fields through multi-wavelength dust polarimetry (see Section \ref{subsec:currentinstru}).

Many detailed reviews such as by \citet{pattle2022} have attempted to put together all the observational and theoretical studies on the roles of magnetic fields in star formation. They have also attempted to answer the questions listed above. In this work, we are presenting a short report on the roles of turbulence and magnetic fields in star formation. Along with global efforts, we have mainly included the efforts of the Indian community (very small though) in answering these questions using facilities in India and abroad. This article consists of Section 2 on models and simulations of the roles of turbulence and B-fields in star formation. Section 3 shows the currently available instruments and global efforts. Section 4 presents the contributions of the Indian community, and Section 5 describes the plans and capabilities for building new facilities and instruments.

\section{Models and simulations}

Two primary reasons for the inefficiency of star formation are turbulence and magnetic fields \citep{FK2012}. The highly supersonic turbulence prevents a global collapse and rearranges the gas in the cloud, forming dense filaments and clumps, which are sites of star formation. This reorganisation of gas affects both the SFR and mass distribution of stars in the cluster (i.e., the initial mass function, IMF). On the other hand, magnetic fields resist the shock compression of gas to high densities and also provide additional magnetic pressure, which suppresses fragmentation, which in turn reduces the SFR \citep{KF2019}. Thus, both turbulence and magnetic fields are crucial components for understanding the physics of star formation. \citet{hocuk2012} studied the impact of B-fields on star formation rate and the IMF. Based on the isothermal B-field simulations, they found that in the presence of B-fields, collapse is slowed down, star formation is found occurring along the field lines, the characteristic mass shifts to smaller mass scale, the log-normal IMF is very bottom heavy, and the IMF has an additional flat high-mass component.

\citet{MF2021} performed numerical simulations of star cluster formation, including a lot of physics such as gravity, turbulence, magnetic fields, and stellar feedback. In this work, authors found that the inclusion of outflows may affect the results by (1) reducing the star-formation rate, (2) increasing fragmentation, and (3) shifting the initial mass function (IMF) to lower masses. These simulations also reproduce the observed mass dependence of multiplicity. The study also finds an average value of specific angular momentum of $\rm 1.5\times 10^{19} cm^{2}~s^{-1}$ which is consistent with observations. In this study, the authors conclude that the IMF is controlled by a combination of gravity, turbulence, magnetic fields, radiation, and outflow feedback. Also, recently, \citet{MFS2022} showed that the star-formation rate and IMF depend on the driving mode (compressive/solenoidal) of the turbulence, which could probably explain the observed lower fraction of low-mass stars in the Galactic Centre \citep{hosek2019}. 

The exact role of turbulence and, especially, magnetic fields in star formation is still an active area of research with numerical efforts from groups around the globe, most recently by \citet{MFS2022}, \citet{GrudicEA2021}, and \cite{BrucyEA2020}. \footnote{Also, see related three-dimensional visualisations of the molecular clouds and stellar feedback from two of these groups: \href{https://www.youtube.com/watch?v=lphp13WPflc&t=1s}{https://www.youtube.com/watch?v=lphp13WPflc\&t=1s} and \href{https://www.youtube.com/watch?v=LeX5e51UkzI}{https://www.youtube.com/watch?v=LeX5e51UkzI}}. These groups are now working towards understanding the effect of the strength and structure of magnetic fields on SFR and IMF to compare the simulated results with theories and observations and gain a better understanding of the star formation process.



\section{Current instruments}\label{subsec:currentinstru}

Polarimetry of interstellar dust is a powerful tool to map interstellar B-fields. There have been major advances in the last few years in polarimetric instrumentation and their use for measuring polarization signals on different scales and density regions of the ISM. These instruments cover the optical, near-, far-IR, and sub(mm) wavelengths. But the most important contribution is put forth by $Planck$ providing the whole sky maps in polarization. Other major contributions in understating the roles of B-fields in star formation are from large surveys on polarization measurements from some ground-based facilities. Some such examples can be the recent results from the James Clerk Maxwell Telescope (JCMT)’s POL-2 polarimeter \citep{friberg2016} on the SCUBA-2 camera \citep{holland2013} operates at 850 $\mu$m and 450 $\mu$m. The JCMT is engaged in a most productive ongoing survey called B-field In STar forming Region Observations (BISTRO; \citealt{ward-thompson2017}). The results from The Atacama Large Millimeter/submillimeter Array (ALMA) \citep{cortes2016} are found very useful in understating B-fields in denser cores. The Atacama Pathfinder Experiment (APEX) telescope is coming up with the PolKa polarimeter at 870$\mu$m \citep{wiesemeyer2014} and the Submillimeter Array (SMA) has an upgraded correlator \citep{primiani2016}. Far-infrared polarization has been detected in many regions from (SOFIA)'s HAWC+ camera \citep{harper2018} and it operates in five bands from 53$\mu$m to 214$\mu$m, while the Balloon-borne Large-Aperture Submillimetre Telescope for Polarimetry (BLASTPol: 250$\mu$m, 350$\mu$m, 500$\mu$m \citep{galitzki2014} and PI-LOT: 214$\mu$m \citep{bernard2016}. Optical and near-IR regimes tracing B-fields in low-density regions have been covered in many studies using SIRPOL on the InfraRed Survey Facility (IRSF; \citet{kandori2006}), Pico dos Dias \citep{magalhaes1996}, the ARIES Imaging Polarimeter (AIMPOL) on the Sampurnanand telescope \citep{rautela2004}, and Mimir on the Perkins Telescope \citep{clemens2007}. A great deal of understanding came into picture by successful survey provided by $Planck$ satellite launched in 2007 with all-sky observations in the sub-mm (at 850 $\mu$m) for understanding magnetic fields at larger scales \citet{planckColVIII2016}.

XPOL -- the Correlation Polarimeter at the IRAM 30-m Telescope \citep{thum2008}.
TolTEC camera of Large Millimeter Telescope (LMT) -- dust continuum polarization at 1.1, 1.4, and 2.0 mm wavelengths at resolutions of 5~--~10$^{"}$ \citep{wilson2020}.
IRAM 30-m telescope of NIKA2pol -- dust continuum polarization at 1.15 mm with a resolution of 11$^{"}$ \citep{ritacco2020}. A list of past, active, and proposed optical to submm polarimeters, is given in table 1.

Radio astronomy is also crucial for observational studies of turbulence and magnetic fields in star-forming galaxies. The interstellar turbulence in the warm/hot phase of the solar neighbourhood, as probed by various radio observations, is consistent with the Kolmogorov theory of turbulence \citep{ArmstrongEA1995}. Based on theory and ultra-high-resolution numerical simulations \citep{FederrathEA2021}, it is known that the turbulence is subsonic or transonic in the warm and hot phases and supersonic in the cold phase. However, observationally, it is difficult to probe turbulence in the ISM. Recent observational studies employing neutral hydrogen spectra in emission show transonic turbulence in the warm (neutral) phase \citep{MarchalEA2021} and a mix of solenoidal and compressive modes of turbulence, both at small \citep{GerradEA2023} and large \citep{NandakumarD2023} scales. The neutral hydrogen spectra in absorption give much higher Mach numbers, which is representative of supersonic turbulence \citep[see Fig. 5 in][]{MurrayEA2015}. These studies utilise high spatial and spectral resolution data from the Green Bank Telescope, Very Large Array, and Australian Square Kilometre Array Pathfinder. 
Similarly, molecular line observations of star-forming cores have revealed a lot on the evolutionary stages of star forming regions. Different molecular species such as CS (2-1), $\rm N_{2}H^{+}$ (1-0), $\rm HCO^{+}$ (1-0), CS(3-2), $\rm DCO^{+}$(2-1) \citep{lee2004} and different transitions of specific molecules trace different chemical and physical conditions and, therefore, can probe different physical scales and densities of the infalling cores. The inward motions in starless cores are one of the essential elements needed to understand the onset of star formation.

\section{Contributions of Indian community}
Over the last one and a half decades, Indian researchers have played an important role in investigating the importance of B-fields, that too, with existing limited facilities. For instance, instruments such as ARIES Imaging POLariemter (AIMPOL) and IUCAA Imaging POLarimeter (IMPOL) are extensively used for mapping polarization in various regimes of ISM such as molecular clouds hosting low- and high-mass star forming regions, HII regions/bubbles, Galactic clusters, dusty filaments, high galactic latitudes etc. These studies led to a good number of publications on magnetic fields \citep{soam2015, soam2019, eswar2019, eswar2020, neha2018, saha2022, ekta2022, bijas2023, sid2023} and dust properties of the ISM using polarization \citep{soam2021}. In these studies, authors have used the multi-band dust polarisation observations mainly from AIMPOL and IMPOL except in some cases where observations are also taken from international facilities. These polarization measurements were used to map the magnetic field morphologies in bright-rimmed clouds (BRCs on the periphery of expanding HII regions), cometary globules (CGs), dark molecular clouds which are potential sites of spontaneous tar formation, and dusty filaments where cores form. This covers a whole regime of isolated and triggered star formation sites in the ISM. Some authors, such as \citet{soam2021}, have used such observations to study dust grain alignment in the ISM. Optical/near-IR polarization observations help invest and test the existing grain alignment theory called Radiative Torque Alignment (RAT). Many studies have used archival polarization data to understand the formation of hub-filament structure (e.g., \citealt{devraj2021}, \citealt{devangan2023}). One of these studies presented magnetic field properties toward the S235 complex using near-infrared (NIR) H-band polarimetric observations, obtained with the Mimir and POLICAN instruments. They found that  S235 complex is a region where stellar feedback triggers new stars, and the magnetic fields regulate the rate of new star formation.

After successful performances of instruments like AIMPOL, there have been some efforts put into developing more polarimeters in India. Recently, instruments such as NICSPol \citep{arthy2019} and EMPOL \citep{ganesh2020} are also developed by the Indian researchers.

 Recently, Indian community have started using Five-hundred-meter Aperture Spherical radio Telescope (FAST) for conducting the H\,I narrow self-absorption (HINSA; \citealt{Li2003}) Zeeman splitting observations towards Planck Galactic Cold Clumps (PGCCs) (Ujwal et al. in prep., Sharma et al. in prep.). Key aim is to quantify the line of sight component of B-field strengths in the clouds. These HINSA Zeeman observations will essentially fill the B-feld strengths in the column density regimes of critical densities where clouds could transform from magnetically supported (sub-critical) to gravitational dominant (super-critical) clouds. A recently published a study on \textit{an early transition to magnetic supercriticality in star formation} by \citet{ching2022} has reported the magnetic field strength of +3.8$\pm$0.3 $\mu$G through HINSA in a prestellar core L1544 \citep{Tafalla1998} using FAST. This core with high central density \citep{Caselli2019} and low central temperature \citep{crapsi2007}, is in an early transition between starless and protostellar phases \citep{Aikawa2001}. Similarly,  Very Large Array (VLA) has produces studies on estimation of core gas temperatures and star formation rates in high redshift Galaxies (VLA-COSMOS; \citealt{Leslie2020}).
In the same line, Giant Metrewave Radio Telescope (GMRT) is also helped in investigating the star-formation rates of external Galaxies \citep{Bera2023}. Now upgraded GMRT (uGMRT) at 685 MHz shows the polarization capabilities which can be used to investigate magnetic fields \citep{Shilpa2021}.


\section{Desired polarimetric capabilities, capacity building, and computing facilities}
In order to progress towards the understanding of B-fields in the star formation processes, it is highly desirable to perform polarization observations at various spatial scales and densities of molecular clouds. We highly desire the following: 

\begin{itemize}
    \item[1.] Currently, India has one fully functional wide-field optical polarimeters (i.e. ARIES IMaging POLarimeter; \citealt{rautela2004}) mounted on 1-m optical telescope in ARIES. Indian Institute of Astrophsics (IIA) host one active photo-polarimeter at their 1-m optical telescope in Kavalur. We do lack spectropolarimeters and infra-red imaging polarimeters. Soam et al. (2024, submitted) are working on developing a spectropolarimeter for a 2-4 m class telescope. The project is still under discussion phase but all the science cases to be addressed from this polarimeter have been included and explained in Soam et al. (2024, submitted). This spectropolarimeter will be covering wavelength range of 0.3–0.7 $\mu$m. The proposed spectral resolution is 100 - 300 with a polarization accuracy ($<\sigma_p>$) of 0.1 and field-of-view of 10 arcmin. Another effort on testing spectropolarimeter at 3.6m telescope in ARIES, Devesthal (Rakshit et al., in prep.), India, is going on. Test observations have been taken and published in \cite{jose2023}. In a parallel work, Srivastava et al. from PRL are developing an echelle spectropolarimeter. With the help of existing polarimetric instrument experts in India and abroad, it is of great importance to plan for developing wide-field polarimeters for 2 -- 4-m class telescopes. An proposal is underway in IIA on making wide-field polarimeter on 2-m telescope.

    \item[2.]  We shall plan to establish a polarimetric-instrumentation group and start discussing by inviting ideas from the polarimetric-instrumentation experts in India and abroad. Community involvement is necessary for such development of facilities and polarimetric instruments.
    
    
    \item[3.] East Asian Observatory (EAO) has been allowing Indian principal investigators (PIs) to submit proposals for JCMT under the ``Expanding Partner Program" -- a program to encourage astronomers from new JCMT partners.  Some of the Indian PIs have already availed observing time on JCMT through this opportunity. More such collaborations/opportunities with other observatories abroad are needed.
    
    We should layout a foundation for India's membership with various observatories across the world to utilize their existing submm facilities. Subsequently, we should also plan for both the single dish and interferometric sub-millimeter telescopes in India  Some discussion is already there on developing a submm facility in India. This will help in conducting spectroscopic, dust continuum, and polarization observations of dense molecular clouds.

    
    \item[4.] Multi-wavelength dust polarization observations along with molecular lines studies can only yield both orientation and strength of the plane-of-the-sky component of magnetic fields. However, in order to fully understand the magnetic field role, a total component of B-field strength is highly desired. In order to measure the line-of-sight component of magnetic field strength, we can plan to carry out both HI and OH Zeeman measurements using a single dish (e.g., FAST, Green Bank telescope) and interferometric facilities (e.g. uGMRT, VLA, etc). uGMRT has already shown its polarization capabilities in \citet{kharb2023, baghel2023}.
    
    \item[5.] To aid the data reduction and analysis of huge amounts of data (e.g. JCMT submm polarimetric data in raw format is $\approx$ 50-60 GB per target, FAST raw data of one target  takes  few to 10's of TB), we may require to establish high-performance computing (HPC) and data storage facilities. Also, high-resolution, state-of-the-art numerical simulations of star formation require millions of CPU hours on supercomputing facilities and these are crucial for comparing simulation results with rich, existing, and upcoming polarimetric observations. Initially we can use some of the existing HPC facilities in the country. Such bilateral efforts would lead to a more detailed and thorough understanding of the role of turbulence and B-fields in star formation.
\end{itemize}


\section{Acknowledgements}

AS thanks Dr.~Archita Rai (KASI, S.~Korea) for the help in getting a list of optical/NIR polarimeters. CE acknowledges the financial support from grant RJF/2020/000071 as a part of the Ramanujan Fellowship awarded by the Science and Engineering Research Board (SERB). ASe thanks Sajay Sunny Mathew, Christoph Federrath, and Antoine Marchal for useful discussions.

\begin{table*}
	\centering
	\caption{Details of the past, current, and proposed polarimeters.}
	\scriptsize
	\begin{tabular}{ccccc}\hline
Polarimeter/Telescope	 & Resolution/FWHM   & wavelength range  & Reference \\    	                                  & (arcsec)                      & $\rm \lambda$ ($\mu$m) &          \\ \hline
  &                          &   UV/Optical/NIR      &         \\ \hline
Polstar              & 30k spectral res.      & UV     & \citet{Paul2022}  \\
AIMPOL/Sampurnanand  &  1.48 arcsec/pixel     & Optical        &   \citet{rautela2004}           \\
EMPOL/Mt. Abu        &  0.18 arcsec/pixel     & Optical        &   \citet{2020SPIE11447E..9EG}   \\
IMPOL/Girawali        &  2 arcsec    & Optical         &   \citet{ram1998}        \\
NICSPol/Mt. Abu        &  0.5 arcsec/pixel     & NIR        &  \citet{Aarthy2019}         \\
NISP/Mt. Abu      &  16.6 arcsec/mm    & NIR        &    \citet{Archita2020}         \\
MFOSC/Mt. Abu      & 15k spectral res.      & 0.4 -0.7        &   \citet{Vipin2022}         \\
MIMIR/Perkins      & 500 spectral res.     & NIR        &   \citet{clemens2007}         \\
SIRPOL/IRSF      &  0.45 arcsec/pixel     & NIR        &     \citet{kandori2006}         \\
AFOSG/Asiago      &  --     & 0.4 -0.7        &     \citet{Pernechele2012}         \\
ALFOSC/NOT      & 6.5 Angstrom/pixel      & Optical & \textcolor{blue}{https://www.not.iac.es/instruments/alfosc/}   \\
FORS2/VLT      &  0.25 arcsec/pixel     & Optical        &   \citet{Gonzalez2020}         \\
WALOP/        &   0.5 arcsec/pixel     & Optical    &   \citet{Maharan2021, Maharana2022}         \\
HOWPol/Kanata        &  --     & optical        & \citet{Kawabata2008}         \\
TFOSC/Russian-Turkish &   0.39 arcsec/pixel  & Optical        &  \citet{Helhel2015}  \\
\hline
 &                    & FIR/submm$^{\dagger}$  &  & \\ 
 \hline
UCL polarimeter  &  300 & 77        &    \citet{cudlip1982}          \\
KAO              &  60  & 270       &    \citet{dragovan1986}         \\
POLY (KAO)       &  55, 40 & 100    &    \citet{novak1989}        \\
MILLIPOL/NRAO 12-m  & 30     & 1300  &   \citet{barvainis1988, clemens1990}  \\
UKT-Pol (JCMT) & 8 --18 & 450, 850, 1100 & \citet{flett1991}  \\
Stokes (Kao)   & 35     & 100             & \citet{Platt1991}       \\
Hertz (CSO)    & 20     & 350        & \citet{schleuning1997, dowell1998}  \\
SCUPOL (JCMT)  & 14     & 850        &  \citet{murray1997, greaves2003}  \\
SPARO VIPER)   & 300    &450         &  \citet{dotson1998, renbarger2004} \\
BLASTPol       & 150     &250, 350, 500 & \citet{galitzki2014} \\
SHARP/SHARC (CSO) & 9    & 350, 450  & \citet{Li2008}\\
POlKa/LABOCA (APEX) & 20  & 870     &  \citet{siringo2012, wiesemeyer2014} \\
HAWC+(SOFIA) & 4.8--18.2 & 53, 62, 89, 154, 214 &  \citet{dowell2018} \\
Planck     & 300   & 850    &  \citet{lamarre2010, planckColVIII2016} \\
PILOT      &120    & 214    &  \citet{foenard2018} \\
POL-2/JCMT &10, 14 & 450, 850 & \citet{bastien2011, friberg2016}\\
BLAST-TNG & 31, 41, 59 & 250, 350, 500 & \citet{galitzki2014} \\
TolTEC (LMT) & 5.0, 6.3, 9.8 &  1100, 1400, 2100 & \citet{bryan2018}\\
NIKA-2/IRAM & 11, 18 & 1150, 2000 & \citet{adam2018} \\
A-MKD (APEX) & 19 (at 850) & 350, 850 &  \citet{otal2014} \\
POL (SPICA) & 9, 18, 32 & 100, 200, 350 & \citet{gaspar2017, roelfsema2018} \\
PICO & 66--192 & 375, 450, 541, 649 & \citet{sutin2018, young2018} \\
FIP (OST)  & 2--10 & 50, 250 & \citet{staguhn2018} \\\hline
	\end{tabular}
	\label{tab:observations}
	
$\dagger$ Information on FIR/submm polarimeters is taken from \citet{pattle2022}.
\end{table*}

\begin{footnotesize}
\bibliography{JAA_main}

\begin{thebibliography}{}
\expandafter\ifx\csname natexlab\endcsname\relax\def\natexlab#1{#1}\fi
\providecommand{\url}[1]{\href{#1}{#1}}
\providecommand{\dodoi}[1]{doi:~\href{http://doi.org/#1}{\nolinkurl{#1}}}
\providecommand{\doeprint}[1]{\href{http://ascl.net/#1}{\nolinkurl{http://ascl.net/#1}}}
\providecommand{\doarXiv}[1]{\href{https://arxiv.org/abs/#1}{\nolinkurl{https://arxiv.org/abs/#1}}}

\bibitem[{{Aarthy} {et~al.}(2019{\natexlab{a}}){Aarthy}, {Rai}, {Ganesh}, \&
  {Vadawale}}]{arthy2019}
{Aarthy}, E., {Rai}, A., {Ganesh}, S., \& {Vadawale}, S.~V. 2019{\natexlab{a}},
  Journal of Astronomical Telescopes, Instruments, and Systems, 5, 035006,
  \dodoi{10.1117/1.JATIS.5.3.035006}

\bibitem[{{Aarthy} {et~al.}(2019{\natexlab{b}}){Aarthy}, {Rai}, {Ganesh}, \&
  {Vadawale}}]{Aarthy2019}
---. 2019{\natexlab{b}}, Journal of Astronomical Telescopes, Instruments, and
  Systems, 5, 035006, \dodoi{10.1117/1.JATIS.5.3.035006}

\bibitem[{{Adam} {et~al.}(2018){Adam}, {Adane}, {Ade}, {Andr{\'e}},
  {Andrianasolo}, {Aussel}, {Beelen}, {Beno{\^\i}t}, {Bideaud}, {Billot},
  {Bourrion}, {Bracco}, {Calvo}, {Catalano}, {Coiffard}, {Comis}, {De Petris},
  {D{\'e}sert}, {Doyle}, {Driessen}, {Evans}, {Goupy}, {Kramer}, {Lagache},
  {Leclercq}, {Leggeri}, {Lestrade}, {Mac{\'\i}as-P{\'e}rez}, {Mauskopf},
  {Mayet}, {Maury}, {Monfardini}, {Navarro}, {Pascale}, {Perotto}, {Pisano},
  {Ponthieu}, {Rev{\'e}ret}, {Rigby}, {Ritacco}, {Romero}, {Roussel}, {Ruppin},
  {Schuster}, {Sievers}, {Triqueneaux}, {Tucker}, \& {Zylka}}]{adam2018}
{Adam}, R., {Adane}, A., {Ade}, P.~A.~R., {et~al.} 2018, \aap, 609, A115,
  \dodoi{10.1051/0004-6361/201731503}

\bibitem[{{Aikawa} {et~al.}(2001){Aikawa}, {Ohashi}, {Inutsuka}, {Herbst}, \&
  {Takakuwa}}]{Aikawa2001}
{Aikawa}, Y., {Ohashi}, N., {Inutsuka}, S.-i., {Herbst}, E., \& {Takakuwa}, S.
  2001, \apj, 552, 639, \dodoi{10.1086/320551}

\bibitem[{{Alves} {et~al.}(2018){Alves}, {Girart}, {Padovani}, {Galli},
  {Franco}, {Caselli}, {Vlemmings}, {Zhang}, \& {Wiesemeyer}}]{Alves2018}
{Alves}, F.~O., {Girart}, J.~M., {Padovani}, M., {et~al.} 2018, \aap, 616, A56,
  \dodoi{10.1051/0004-6361/201832935}

\bibitem[{{Anderson} {et~al.}(2014){Anderson}, {Bania}, {Balser}, {Cunningham},
  {Wenger}, {Johnstone}, \& {Armentrout}}]{anderson2014}
{Anderson}, L.~D., {Bania}, T.~M., {Balser}, D.~S., {et~al.} 2014, \apjs, 212,
  1, \dodoi{10.1088/0067-0049/212/1/1}

\bibitem[{{Anderson} {et~al.}(2011){Anderson}, {Bania}, {Balser}, \&
  {Rood}}]{anderson2011}
{Anderson}, L.~D., {Bania}, T.~M., {Balser}, D.~S., \& {Rood}, R.~T. 2011,
  \apjs, 194, 32, \dodoi{10.1088/0067-0049/194/2/32}

\bibitem[{{Armstrong} {et~al.}(1995){Armstrong}, {Rickett}, \&
  {Spangler}}]{ArmstrongEA1995}
{Armstrong}, J.~W., {Rickett}, B.~J., \& {Spangler}, S.~R. 1995, \apj, 443,
  209, \dodoi{10.1086/175515}

\bibitem[{{Baghel} {et~al.}(2023){Baghel}, {Kharb}, {Silpa}, {Ho}, \&
  {Harrison}}]{baghel2023}
{Baghel}, J., {Kharb}, P., {Silpa}, S., {Ho}, L.~C., \& {Harrison}, C.~M. 2023,
  arXiv e-prints, arXiv:2304.11831, \dodoi{10.48550/arXiv.2304.11831}

\bibitem[{{Barvainis} {et~al.}(1988){Barvainis}, {Clemens}, \&
  {Leach}}]{barvainis1988}
{Barvainis}, R., {Clemens}, D.~P., \& {Leach}, R. 1988, \aj, 95, 510,
  \dodoi{10.1086/114650}

\bibitem[{{Bastien} {et~al.}(2011){Bastien}, {Bissonnette}, {Simon},
  {Coud{\'e}}, {Ade}, {Savini}, {Pisano}, {Leclerc}, {Bernier}, {Landry},
  {Houde}, {Hezareh}, {Naylor}, {Gom}, {Jenness}, {Berry}, {Johnstone}, \&
  {Matthews}}]{bastien2011}
{Bastien}, P., {Bissonnette}, E., {Simon}, A., {et~al.} 2011, in Astronomical
  Society of the Pacific Conference Series, Vol. 449, Astronomical Polarimetry
  2008: Science from Small to Large Telescopes, ed. P.~{Bastien}, N.~{Manset},
  D.~P. {Clemens}, \& N.~{St-Louis}, 68

\bibitem[{{Beltr{\'a}n} {et~al.}(2019){Beltr{\'a}n}, {Padovani}, {Girart},
  {Galli}, {Cesaroni}, {Paladino}, {Anglada}, {Estalella}, {Osorio}, {Rao},
  {S{\'a}nchez-Monge}, \& {Zhang}}]{Beltran2019}
{Beltr{\'a}n}, M.~T., {Padovani}, M., {Girart}, J.~M., {et~al.} 2019, \aap,
  630, A54, \dodoi{10.1051/0004-6361/201935701}

\bibitem[{{Benson} \& {Myers}(1989)}]{benson&Myers1989}
{Benson}, P.~J., \& {Myers}, P.~C. 1989, \apjs, 71, 89, \dodoi{10.1086/191365}

\bibitem[{{Bera} {et~al.}(2023){Bera}, {Kanekar}, {Chengalur}, \&
  {Bagla}}]{Bera2023}
{Bera}, A., {Kanekar}, N., {Chengalur}, J.~N., \& {Bagla}, J.~S. 2023, \apjl,
  956, L15, \dodoi{10.3847/2041-8213/acf71a}

\bibitem[{{Bernard} {et~al.}(2016){Bernard}, {Ade}, {Andr{\'e}}, {Aumont},
  {Bautista}, {Bray}, {Bernardis}, {Boulade}, {Bousquet}, {Bouzit}, {Buttice},
  {Caillat}, {Charra}, {Chaigneau}, {Crane}, {Crussaire}, {Douchin},
  {Doumayrou}, {Dubois}, {Engel}, {Etcheto}, {G{\'e}lot}, {Griffin}, {Foenard},
  {Grabarnik}, {Hargrave}, {Hughes}, {Laureijs}, {Lepennec}, {Leriche},
  {Longval}, {Maestre}, {Maffei}, {Martignac}, {Marty}, {Marty}, {Masi},
  {Mirc}, {Misawa}, {Montel}, {Montier}, {Mot}, {Narbonne}, {Nicot}, {Pajot},
  {Parot}, {P{\'e}rot}, {Pimentao}, {Pisano}, {Ponthieu}, {Ristorcelli},
  {Rodriguez}, {Roudil}, {Salatino}, {Savini}, {Simonella}, {Saccoccio},
  {Tapie}, {Tauber}, {Torre}, \& {Tucker}}]{bernard2016}
{Bernard}, J.~P., {Ade}, P., {Andr{\'e}}, Y., {et~al.} 2016, Experimental
  Astronomy, 42, 199, \dodoi{10.1007/s10686-016-9506-1}

\bibitem[{{Bijas} {et~al.}(2023){Bijas}, {Eswaraiah}, {Wang}, {Jose}, {Chen},
  {Li}, {Lai}, \& {Ojha}}]{bijas2023}
{Bijas}, N., {Eswaraiah}, C., {Wang}, J.-W., {et~al.} 2023, \mnras, 526, 1308,
  \dodoi{10.1093/mnras/stad2880}

\bibitem[{{Borlaff} {et~al.}(2021){Borlaff}, {Lopez-Rodriguez}, {Beck},
  {Stepanov}, {Ntormousi}, {Hughes}, {Tassis}, {Marcum}, {Grosset}, {Beckman},
  {Proudfit}, {Clark}, {D{\'\i}az-Santos}, {Mao}, {Reach}, {Roman-Duval},
  {Subramanian}, {Tram}, {Zweibel}, {Dale}, \& {Legacy Team}}]{BorlaffEA2021}
{Borlaff}, A.~S., {Lopez-Rodriguez}, E., {Beck}, R., {et~al.} 2021, \apj, 921,
  128, \dodoi{10.3847/1538-4357/ac16d7}

\bibitem[{{Bracco} {et~al.}(2020){Bracco}, {Jeli{\'c}}, {Marchal}, {Turi{\'c}},
  {Erceg}, {Miville-Desch{\^e}nes}, \& {Bellomi}}]{BraccoEA2020}
{Bracco}, A., {Jeli{\'c}}, V., {Marchal}, A., {et~al.} 2020, \aap, 644, L3,
  \dodoi{10.1051/0004-6361/202039283}

\bibitem[{{Brucy} {et~al.}(2020){Brucy}, {Hennebelle}, {Bournaud}, \&
  {Colling}}]{BrucyEA2020}
{Brucy}, N., {Hennebelle}, P., {Bournaud}, F., \& {Colling}, C. 2020, \apjl,
  896, L34, \dodoi{10.3847/2041-8213/ab9830}

\bibitem[{{Bryan}(2018)}]{bryan2018}
{Bryan}, S. 2018, in Atacama Large-Aperture Submm/mm Telescope (AtLAST), 36,
  \dodoi{10.5281/zenodo.1159073}

\bibitem[{{Caselli} {et~al.}(2019){Caselli}, {Pineda}, {Zhao}, {Walmsley},
  {Keto}, {Tafalla}, {Chac{\'o}n-Tanarro}, {Bourke}, {Friesen}, {Galli}, \&
  {Padovani}}]{Caselli2019}
{Caselli}, P., {Pineda}, J.~E., {Zhao}, B., {et~al.} 2019, \apj, 874, 89,
  \dodoi{10.3847/1538-4357/ab0700}

\bibitem[{{Ching} {et~al.}(2022){Ching}, {Li}, {Heiles}, {Li}, {Qian}, {Yue},
  {Tang}, \& {Jiao}}]{ching2022}
{Ching}, T.~C., {Li}, D., {Heiles}, C., {et~al.} 2022, \nat, 601, 49,
  \dodoi{10.1038/s41586-021-04159-x}

\bibitem[{{Clemens} {et~al.}(1990){Clemens}, {Leach}, {Barvainis}, \&
  {Kane}}]{clemens1990}
{Clemens}, D.~P., {Leach}, R.~W., {Barvainis}, R., \& {Kane}, B.~D. 1990,
  \pasp, 102, 1064, \dodoi{10.1086/132735}

\bibitem[{{Clemens} {et~al.}(2007){Clemens}, {Sarcia}, {Grabau}, {Tollestrup},
  {Buie}, {Dunham}, \& {Taylor}}]{clemens2007}
{Clemens}, D.~P., {Sarcia}, D., {Grabau}, A., {et~al.} 2007, \pasp, 119, 1385,
  \dodoi{10.1086/524775}

\bibitem[{{Cortes} {et~al.}(2016){Cortes}, {Girart}, {Hull}, {Sridharan},
  {Louvet}, {Plambeck}, {Li}, {Crutcher}, \& {Lai}}]{cortes2016}
{Cortes}, P.~C., {Girart}, J.~M., {Hull}, C. L.~H., {et~al.} 2016, \apjl, 825,
  L15, \dodoi{10.3847/2041-8205/825/1/L15}

\bibitem[{{Crapsi} {et~al.}(2007){Crapsi}, {Caselli}, {Walmsley}, \&
  {Tafalla}}]{crapsi2007}
{Crapsi}, A., {Caselli}, P., {Walmsley}, M.~C., \& {Tafalla}, M. 2007, \aap,
  470, 221, \dodoi{10.1051/0004-6361:20077613}

\bibitem[{{Crutcher} \& {Kemball}(2019)}]{Crutcher&Richard2019}
{Crutcher}, R.~M., \& {Kemball}, A.~J. 2019, Frontiers in Astronomy and Space
  Sciences, 6, 66, \dodoi{10.3389/fspas.2019.00066}

\bibitem[{{Cudlip} {et~al.}(1982){Cudlip}, {Furniss}, {King}, \&
  {Jennings}}]{cudlip1982}
{Cudlip}, W., {Furniss}, I., {King}, K.~J., \& {Jennings}, R.~E. 1982, \mnras,
  200, 1169, \dodoi{10.1093/mnras/200.4.1169}

\bibitem[{{Dale} {et~al.}(2014){Dale}, {Ngoumou}, {Ercolano}, \&
  {Bonnell}}]{dale2014}
{Dale}, J.~E., {Ngoumou}, J., {Ercolano}, B., \& {Bonnell}, I.~A. 2014, \mnras,
  442, 694, \dodoi{10.1093/mnras/stu816}

\bibitem[{{Dall'Olio} {et~al.}(2019){Dall'Olio}, {Vlemmings}, {Persson},
  {Alves}, {Beuther}, {Girart}, {Surcis}, {Torrelles}, \& {Van
  Langevelde}}]{Dall'Olio2019}
{Dall'Olio}, D., {Vlemmings}, W.~H.~T., {Persson}, M.~V., {et~al.} 2019, \aap,
  626, A36, \dodoi{10.1051/0004-6361/201834100}

\bibitem[{{Devaraj} {et~al.}(2021){Devaraj}, {Clemens}, {Dewangan}, {Luna},
  {Ray}, \& {Mackey}}]{devraj2021}
{Devaraj}, R., {Clemens}, D.~P., {Dewangan}, L.~K., {et~al.} 2021, \apj, 911,
  81, \dodoi{10.3847/1538-4357/abe9b1}

\bibitem[{{Dewangan} {et~al.}(2023){Dewangan}, {Bhadari}, {Men'shchikov},
  {Chung}, {Devaraj}, {Lee}, {Maity}, \& {Baug}}]{devangan2023}
{Dewangan}, L.~K., {Bhadari}, N.~K., {Men'shchikov}, A., {et~al.} 2023, \apj,
  946, 22, \dodoi{10.3847/1538-4357/acbccc}

\bibitem[{{Dotson} {et~al.}(1998){Dotson}, {Novak}, {Renbarger}, {Pernic}, \&
  {Sundwall}}]{dotson1998}
{Dotson}, J.~L., {Novak}, G., {Renbarger}, T., {Pernic}, D., \& {Sundwall},
  J.~L. 1998, in Society of Photo-Optical Instrumentation Engineers (SPIE)
  Conference Series, Vol. 3357, Advanced Technology MMW, Radio, and Terahertz
  Telescopes, ed. T.~G. {Phillips}, 543--547, \dodoi{10.1117/12.317388}

\bibitem[{{Dowell} {et~al.}(2018){Dowell}, {HAWC+ Instrument Team}, \& {HAWC+
  Science Team}}]{dowell2018}
{Dowell}, C.~D., {HAWC+ Instrument Team}, \& {HAWC+ Science Team}. 2018, in
  American Astronomical Society Meeting Abstracts, Vol. 232, American
  Astronomical Society Meeting Abstracts \#232, 103.05

\bibitem[{{Dowell} {et~al.}(1998){Dowell}, {Hildebrand}, {Schleuning},
  {Vaillancourt}, {Dotson}, {Novak}, {Renbarger}, \& {Houde}}]{dowell1998}
{Dowell}, C.~D., {Hildebrand}, R.~H., {Schleuning}, D.~A., {et~al.} 1998, \apj,
  504, 588, \dodoi{10.1086/306069}

\bibitem[{{Dragovan}(1986)}]{dragovan1986}
{Dragovan}, M. 1986, PhD thesis, University of Chicago

\bibitem[{{Elmegreen} \& {Scalo}(2004)}]{elmegreen2004}
{Elmegreen}, B.~G., \& {Scalo}, J. 2004, \araa, 42, 211,
  \dodoi{10.1146/annurev.astro.41.011802.094859}

\bibitem[{{Eswaraiah} {et~al.}(2017){Eswaraiah}, {Lai}, {Chen}, {Pandey},
  {Tamura}, {Maheswar}, {Sharma}, {Wang}, {Nishiyama}, {Nakajima}, {Kwon},
  {Purcell}, \& {Magalh{\~a}es}}]{eswar2017}
{Eswaraiah}, C., {Lai}, S.-P., {Chen}, W.-P., {et~al.} 2017, \apj, 850, 195,
  \dodoi{10.3847/1538-4357/aa917e}

\bibitem[{{Eswaraiah} {et~al.}(2019){Eswaraiah}, {Lai}, {Ma}, {Pandey}, {Jose},
  {Chen}, {Samal}, {Wang}, {Sharma}, \& {Ojha}}]{eswar2019}
{Eswaraiah}, C., {Lai}, S.-P., {Ma}, Y., {et~al.} 2019, \apj, 875, 64,
  \dodoi{10.3847/1538-4357/ab0a0c}

\bibitem[{{Eswaraiah} {et~al.}(2020){Eswaraiah}, {Li}, {Samal}, {Wang}, {Ma},
  {Lai}, {Zavagno}, {Ching}, {Liu}, {Pattle}, {Ward-Thompson}, {Pandey}, \&
  {Ojha}}]{eswar2020}
{Eswaraiah}, C., {Li}, D., {Samal}, M.~R., {et~al.} 2020, \apj, 897, 90,
  \dodoi{10.3847/1538-4357/ab83f2}

\bibitem[{{Eswaraiah} {et~al.}(2021){Eswaraiah}, {Li}, {Furuya}, {Hasegawa},
  {Ward-Thompson}, {Qiu}, {Ohashi}, {Pattle}, {Sadavoy}, {Hull}, {Berry},
  {Doi}, {Ching}, {Lai}, {Wang}, {Koch}, {Kwon}, {Kwon}, {Bastien},
  {Arzoumanian}, {Coud{\'e}}, {Soam}, {Fanciullo}, {Yen}, {Liu}, {Hoang}, {Ping
  Chen}, {Shimajiri}, {Liu}, {Chen}, {Li}, {Lyo}, {Hwang}, {Johnstone}, {Rao},
  {Bich Ngoc}, {Ngoc Diep}, {Mairs}, {Parsons}, {Tamura}, {Tahani}, {Vivien
  Chen}, {Nakamura}, {Shinnaga}, {Tang}, {Cho}, {Won Lee}, {Inutsuka}, {Inoue},
  {Iwasaki}, {Qian}, {Xie}, {Li}, {Liu}, {Zhang}, {Chen}, {Zhang}, {Zhu},
  {Zhou}, {Andr{\'e}}, {Liu}, {Yuan}, {Lu}, {Peretto}, {Bourke}, {Byun}, {Dai},
  {Duan}, {Duan}, {Eden}, {Matthews}, {Fiege}, {Fissel}, {Kim}, {Lee}, {Kim},
  {Pyo}, {Choi}, {Choi}, {Chrysostomou}, {Jung Chung}, {Ngoc Tram},
  {Franzmann}, {Friberg}, {Friesen}, {Fuller}, {Gledhill}, {Graves}, {Greaves},
  {Griffin}, {Gu}, {Han}, {Hatchell}, {Hayashi}, {Houde}, {Kawabata}, {Jeong},
  {Kang}, {Kang}, {Kang}, {Kataoka}, {Kemper}, {Rawlings}, {Rawlings},
  {Retter}, {Richer}, {Rigby}, {Saito}, {Savini}, {Scaife}, {Seta}, {Kim}, {Hee
  Kim}, {Kim}, {Kirchschlager}, {Kirk}, {Kobayashi}, {Konyves}, {Kusune},
  {Lacaille}, {Law}, {Lee}, {Lee}, {Matsumura}, {Moriarty-Schieven}, {Nagata},
  {Nakanishi}, {Onaka}, {Park}, {Tang}, {Tomisaka}, {Tsukamoto}, {Viti},
  {Wang}, {Whitworth}, {Yoo}, {Yun}, {Zenko}, {Zhang}, {de Looze}, {Dowell},
  {Eyres}, {Falle}, {Robitaille}, \& {van Loo}}]{eswar2021}
{Eswaraiah}, C., {Li}, D., {Furuya}, R.~S., {et~al.} 2021, \apjl, 912, L27,
  \dodoi{10.3847/2041-8213/abeb1c}

\bibitem[{{Federrath} \& {Klessen}(2012)}]{FK2012}
{Federrath}, C., \& {Klessen}, R.~S. 2012, \apj, 761, 156,
  \dodoi{10.1088/0004-637X/761/2/156}

\bibitem[{{Federrath} {et~al.}(2021){Federrath}, {Klessen}, {Iapichino}, \&
  {Beattie}}]{FederrathEA2021}
{Federrath}, C., {Klessen}, R.~S., {Iapichino}, L., \& {Beattie}, J.~R. 2021,
  Nature Astronomy, 5, 365, \dodoi{10.1038/s41550-020-01282-z}

\bibitem[{{Ferri{\`e}re}(2020)}]{fer2020}
{Ferri{\`e}re}, K. 2020, Plasma Physics and Controlled Fusion, 62, 014014,
  \dodoi{10.1088/1361-6587/ab49eb}

\bibitem[{{Flett} \& {Murray}(1991)}]{flett1991}
{Flett}, A.~M., \& {Murray}, A.~G. 1991, \mnras, 249, 4P,
  \dodoi{10.1093/mnras/249.1.4P}

\bibitem[{{Friberg} {et~al.}(2016){Friberg}, {Bastien}, {Berry}, {Savini},
  {Graves}, \& {Pattle}}]{friberg2016}
{Friberg}, P., {Bastien}, P., {Berry}, D., {et~al.} 2016, in Society of
  Photo-Optical Instrumentation Engineers (SPIE) Conference Series, Vol. 9914,
  Millimeter, Submillimeter, and Far-Infrared Detectors and Instrumentation for
  Astronomy VIII, ed. W.~S. {Holland} \& J.~{Zmuidzinas}, 991403,
  \dodoi{10.1117/12.2231943}

\bibitem[{{Galitzki} {et~al.}(2014){Galitzki}, {Ade}, {Angil{\`e}}, {Benton},
  {Devlin}, {Dober}, {Fissel}, {Fukui}, {Gandilo}, {Klein}, {Korotkov},
  {Matthews}, {Moncelsi}, {Netterfield}, {Novak}, {Nutter}, {Pascale},
  {Poidevin}, {Savini}, {Scott}, {Shariff}, {Soler}, {Tucker}, {Tucker}, \&
  {Ward-Thompson}}]{galitzki2014}
{Galitzki}, N., {Ade}, P.~A.~R., {Angil{\`e}}, F.~E., {et~al.} 2014, in Society
  of Photo-Optical Instrumentation Engineers (SPIE) Conference Series, Vol.
  9145, Ground-based and Airborne Telescopes V, ed. L.~M. {Stepp},
  R.~{Gilmozzi}, \& H.~J. {Hall}, 91450R, \dodoi{10.1117/12.2054759}

\bibitem[{{Ganesh} {et~al.}(2020{\natexlab{a}}){Ganesh}, {Rai}, {Aravind},
  {Singh}, {Prajapati}, {Mishra}, {Kasarla}, {Sarkar}, {Patwal}, {Uppal},
  {Chandra}, {Mathur}, {Shah}, {Baliyan}, \& {Joshi}}]{ganesh2020}
{Ganesh}, S., {Rai}, A., {Aravind}, K., {et~al.} 2020{\natexlab{a}}, in Society
  of Photo-Optical Instrumentation Engineers (SPIE) Conference Series, Vol.
  11447, Ground-based and Airborne Instrumentation for Astronomy VIII, ed.
  C.~J. {Evans}, J.~J. {Bryant}, \& K.~{Motohara}, 114479E,
  \dodoi{10.1117/12.2560949}

\bibitem[{{Ganesh} {et~al.}(2020{\natexlab{b}}){Ganesh}, {Rai}, {Aravind},
  {Singh}, {Prajapati}, {Mishra}, {Kasarla}, {Sarkar}, {Patwal}, {Uppal},
  {Chandra}, {Mathur}, {Shah}, {Baliyan}, \& {Joshi}}]{2020SPIE11447E..9EG}
{Ganesh}, S., {Rai}, A., {Aravind}, K., {et~al.} 2020{\natexlab{b}}, in Society
  of Photo-Optical Instrumentation Engineers (SPIE) Conference Series, Vol.
  11447, Ground-based and Airborne Instrumentation for Astronomy VIII, ed.
  C.~J. {Evans}, J.~J. {Bryant}, \& K.~{Motohara}, 114479E,
  \dodoi{10.1117/12.2560949}

\bibitem[{{Gaspar Venancio} {et~al.}(2017){Gaspar Venancio}, {Doyle}, {Isaak},
  {Onaka}, {Kaneda}, {Nakagawa}, {Matsuhara}, {Takahashi}, \&
  {Yamawaki}}]{gaspar2017}
{Gaspar Venancio}, L.~M., {Doyle}, D., {Isaak}, K., {et~al.} 2017, in Society
  of Photo-Optical Instrumentation Engineers (SPIE) Conference Series, Vol.
  10565, Society of Photo-Optical Instrumentation Engineers (SPIE) Conference
  Series, 1056555, \dodoi{10.1117/12.2309162}

\bibitem[{{Geen} {et~al.}(2020){Geen}, {Pellegrini}, {Bieri}, \&
  {Klessen}}]{geen2020}
{Geen}, S., {Pellegrini}, E., {Bieri}, R., \& {Klessen}, R. 2020, \mnras, 492,
  915, \dodoi{10.1093/mnras/stz3491}

\bibitem[{{Gent} {et~al.}(2023){Gent}, {Mac Low}, {Korpi-Lagg}, \&
  {Singh}}]{GentEA2023}
{Gent}, F.~A., {Mac Low}, M.-M., {Korpi-Lagg}, M.~J., \& {Singh}, N.~K. 2023,
  \apj, 943, 176, \dodoi{10.3847/1538-4357/acac20}

\bibitem[{{Gerrard} {et~al.}(2023){Gerrard}, {Federrath}, {Pingel},
  {McClure-Griffiths}, {Marchal}, {Joncas}, {Clark}, {Stanimirovi{\'c}}, {Lee},
  {van Loon}, {Dickey}, {D{\'e}nes}, {Ma}, {Dempsey}, \& {Lynn}}]{GerradEA2023}
{Gerrard}, I.~A., {Federrath}, C., {Pingel}, N.~M., {et~al.} 2023, \mnras, 526,
  982, \dodoi{10.1093/mnras/stad2718}

\bibitem[{{Girart} {et~al.}(2018){Girart}, {Fern{\'a}ndez-L{\'o}pez}, {Li},
  {Yang}, {Estalella}, {Anglada}, {{\'A}{\~n}ez-L{\'o}pez}, {Busquet},
  {Carrasco-Gonz{\'a}lez}, {Curiel}, {Galvan-Madrid}, {G{\'o}mez}, {de
  Gregorio-Monsalvo}, {Jim{\'e}nez-Serra}, {Krasnopolsky}, {Mart{\'\i}},
  {Osorio}, {Padovani}, {Rao}, {Rodr{\'\i}guez}, \& {Torrelles}}]{Girart2018}
{Girart}, J.~M., {Fern{\'a}ndez-L{\'o}pez}, M., {Li}, Z.~Y., {et~al.} 2018,
  \apjl, 856, L27, \dodoi{10.3847/2041-8213/aab76b}

\bibitem[{{Gonz{\'a}lez-Gait{\'a}n} {et~al.}(2020){Gonz{\'a}lez-Gait{\'a}n},
  {Mour{\~a}o}, {Patat}, {Anderson}, {Cikota}, {Wiersema}, {Higgins}, \&
  {Silva}}]{Gonzalez2020}
{Gonz{\'a}lez-Gait{\'a}n}, S., {Mour{\~a}o}, A.~M., {Patat}, F., {et~al.} 2020,
  \aap, 634, A70, \dodoi{10.1051/0004-6361/201936379}

\bibitem[{{Greaves} {et~al.}(2003){Greaves}, {Holland}, {Jenness},
  {Chrysostomou}, {Berry}, {Murray}, {Tamura}, {Robson}, {Ade}, {Nartallo},
  {Stevens}, {Momose}, {Morino}, {Moriarty-Schieven}, {Gannaway}, \&
  {Haynes}}]{greaves2003}
{Greaves}, J.~S., {Holland}, W.~S., {Jenness}, T., {et~al.} 2003, \mnras, 340,
  353, \dodoi{10.1046/j.1365-8711.2003.06230.x}

\bibitem[{{Grudi{\'c}} {et~al.}(2021){Grudi{\'c}}, {Guszejnov}, {Hopkins},
  {Offner}, \& {Faucher-Gigu{\`e}re}}]{GrudicEA2021}
{Grudi{\'c}}, M.~Y., {Guszejnov}, D., {Hopkins}, P.~F., {Offner}, S. S.~R., \&
  {Faucher-Gigu{\`e}re}, C.-A. 2021, \mnras, 506, 2199,
  \dodoi{10.1093/mnras/stab1347}

\bibitem[{{Harper} {et~al.}(2018){Harper}, {Runyan}, {Dowell}, {Wirth},
  {Amato}, {Ames}, {Amiri}, {Banks}, {Bartels}, {Benford}, {Berthoud},
  {Buchanan}, {Casey}, {Chapman}, {Chuss}, {Cook}, {Derro}, {Dotson}, {Evans},
  {Fixsen}, {Gatley}, {Guerra}, {Halpern}, {Hamilton}, {Hamlin}, {Hansen},
  {Heimsath}, {Hermida}, {Hilton}, {Hirsch}, {Hollister}, {Hostetter}, {Irwin},
  {Jhabvala}, {Jhabvala}, {Kastner}, {Kov{\'a}cs}, {Lin}, {Loewenstein},
  {Looney}, {Lopez-Rodriguez}, {Maher}, {Michail}, {Miller}, {Moseley},
  {Novak}, {Pernic}, {Rennick}, {Rhody}, {Sandberg}, {Sandford}, {Santos},
  {Shafer}, {Sharp}, {Shirron}, {Siah}, {Silverberg}, {Sparr}, {Spotz},
  {Staguhn}, {Toorian}, {Towey}, {Tuttle}, {Vaillancourt}, {Voellmer},
  {Volpert}, {Wang}, \& {Wollack}}]{harper2018}
{Harper}, D.~A., {Runyan}, M.~C., {Dowell}, C.~D., {et~al.} 2018, Journal of
  Astronomical Instrumentation, 7, 1840008, \dodoi{10.1142/S2251171718400081}

\bibitem[{{Hayward} \& {Hopkins}(2017)}]{Hayward2017}
{Hayward}, C.~C., \& {Hopkins}, P.~F. 2017, \mnras, 465, 1682,
  \dodoi{10.1093/mnras/stw2888}

\bibitem[{{Helhel} {et~al.}(2015){Helhel}, {Khamitov}, {Kahya}, {Bayar},
  {Kaynar}, \& {Gumerov}}]{Helhel2015}
{Helhel}, S., {Khamitov}, I., {Kahya}, G., {et~al.} 2015, Experimental
  Astronomy, 39, 595, \dodoi{10.1007/s10686-015-9468-8}

\bibitem[{{Hennebelle} \& {Chabrier}(2008)}]{hennebelle2008}
{Hennebelle}, P., \& {Chabrier}, G. 2008, \apj, 684, 395,
  \dodoi{10.1086/589916}

\bibitem[{{Hocuk} {et~al.}(2012){Hocuk}, {Schleicher}, {Spaans}, \&
  {Cazaux}}]{hocuk2012}
{Hocuk}, S., {Schleicher}, D.~R.~G., {Spaans}, M., \& {Cazaux}, S. 2012, \aap,
  545, A46, \dodoi{10.1051/0004-6361/201219628}

\bibitem[{{Holland} {et~al.}(2013){Holland}, {Bintley}, {Chapin},
  {Chrysostomou}, {Davis}, {Dempsey}, {Duncan}, {Fich}, {Friberg}, {Halpern},
  {Irwin}, {Jenness}, {Kelly}, {MacIntosh}, {Robson}, {Scott}, {Ade},
  {Atad-Ettedgui}, {Berry}, {Craig}, {Gao}, {Gibb}, {Hilton}, {Hollister},
  {Kycia}, {Lunney}, {McGregor}, {Montgomery}, {Parkes}, {Tilanus}, {Ullom},
  {Walther}, {Walton}, {Woodcraft}, {Amiri}, {Atkinson}, {Burger}, {Chuter},
  {Coulson}, {Doriese}, {Dunare}, {Economou}, {Niemack}, {Parsons},
  {Reintsema}, {Sibthorpe}, {Smail}, {Sudiwala}, \& {Thomas}}]{holland2013}
{Holland}, W.~S., {Bintley}, D., {Chapin}, E.~L., {et~al.} 2013, \mnras, 430,
  2513, \dodoi{10.1093/mnras/sts612}

\bibitem[{{Hosek} {et~al.}(2019){Hosek}, {Lu}, {Anderson}, {Najarro}, {Ghez},
  {Morris}, {Clarkson}, \& {Albers}}]{hosek2019}
{Hosek}, Matthew~W., J., {Lu}, J.~R., {Anderson}, J., {et~al.} 2019, \apj, 870,
  44, \dodoi{10.3847/1538-4357/aaef90}

\bibitem[{{Jenkins} \& {Tripp}(2011)}]{JenkinsTripp2011}
{Jenkins}, E.~B., \& {Tripp}, T.~M. 2011, \apj, 734, 65,
  \dodoi{10.1088/0004-637X/734/1/65}

\bibitem[{{Jose} {et~al.}(2023){Jose}, {Rakshit}, {Pandey}, \&
  {Omar}}]{jose2023}
{Jose}, J., {Rakshit}, S., {Pandey}, S., \& {Omar}, A. 2023, The Astronomer's
  Telegram, 15931, 1

\bibitem[{{Kandori} {et~al.}(2006){Kandori}, {Kusakabe}, {Tamura}, {Nakajima},
  {Nagayama}, {Nagashima}, {Hashimoto}, {Hough}, {Sato}, {Nagata}, {Ishihara},
  {Lucas}, \& {Fukagawa}}]{kandori2006}
{Kandori}, R., {Kusakabe}, N., {Tamura}, M., {et~al.} 2006, in Society of
  Photo-Optical Instrumentation Engineers (SPIE) Conference Series, Vol. 6269,
  Society of Photo-Optical Instrumentation Engineers (SPIE) Conference Series,
  ed. I.~S. {McLean} \& M.~{Iye}, 626951, \dodoi{10.1117/12.670967}

\bibitem[{{Kawabata} {et~al.}(2008){Kawabata}, {Nagae}, {Chiyonobu}, {Tanaka},
  {Nakaya}, {Suzuki}, {Kamata}, {Miyazaki}, {Hiragi}, {Miyamoto}, {Yamanaka},
  {Arai}, {Yamashita}, {Uemura}, {Ohsugi}, {Isogai}, {Ishitobi}, \&
  {Sato}}]{Kawabata2008}
{Kawabata}, K.~S., {Nagae}, O., {Chiyonobu}, S., {et~al.} 2008, in Society of
  Photo-Optical Instrumentation Engineers (SPIE) Conference Series, Vol. 7014,
  Ground-based and Airborne Instrumentation for Astronomy II, ed. I.~S.
  {McLean} \& M.~M. {Casali}, 70144L, \dodoi{10.1117/12.788569}

\bibitem[{{Kharb} {et~al.}(2023){Kharb}, {Sasikumar}, {Baghel}, \&
  {Ghosh}}]{kharb2023}
{Kharb}, P., {Sasikumar}, S., {Baghel}, J., \& {Ghosh}, S. 2023, arXiv
  e-prints, arXiv:2305.04420, \dodoi{10.48550/arXiv.2305.04420}

\bibitem[{{Kim} {et~al.}(2021){Kim}, {Ostriker}, \& {Filippova}}]{kim2021}
{Kim}, J.-G., {Ostriker}, E.~C., \& {Filippova}, N. 2021, \apj, 911, 128,
  \dodoi{10.3847/1538-4357/abe934}

\bibitem[{{Krumholz}(2014)}]{krumholz2014}
{Krumholz}, M.~R. 2014, \physrep, 539, 49,
  \dodoi{10.1016/j.physrep.2014.02.001}

\bibitem[{{Krumholz} {et~al.}(2018){Krumholz}, {Burkhart}, {Forbes}, \&
  {Crocker}}]{KEA2018}
{Krumholz}, M.~R., {Burkhart}, B., {Forbes}, J.~C., \& {Crocker}, R.~M. 2018,
  \mnras, 477, 2716, \dodoi{10.1093/mnras/sty852}

\bibitem[{{Krumholz} \& {Federrath}(2019)}]{KF2019}
{Krumholz}, M.~R., \& {Federrath}, C. 2019, Frontiers in Astronomy and Space
  Sciences, 6, 7, \dodoi{10.3389/fspas.2019.00007}

\bibitem[{{Krumholz} {et~al.}(2005){Krumholz}, {Klein}, \&
  {McKee}}]{krumholz2005}
{Krumholz}, M.~R., {Klein}, R.~I., \& {McKee}, C.~F. 2005, in Protostars and
  Planets V Posters, 8271

\bibitem[{{Krumholz} \& {Tan}(2007)}]{KT2007}
{Krumholz}, M.~R., \& {Tan}, J.~C. 2007, \apj, 654, 304, \dodoi{10.1086/509101}

\bibitem[{{Kumar} {et~al.}(2023){Kumar}, {Soam}, \& {Roy}}]{sid2023}
{Kumar}, S., {Soam}, A., \& {Roy}, N. 2023, \mnras, 524, 1219,
  \dodoi{10.1093/mnras/stad1845}

\bibitem[{{Kumar} {et~al.}(2022){Kumar}, {Srivastava}, {Dixit}, {Mistry},
  {Lad}, {Patel}, \& {Rajpurohit}}]{Vipin2022}
{Kumar}, V., {Srivastava}, M.~K., {Dixit}, V., {et~al.} 2022, in Society of
  Photo-Optical Instrumentation Engineers (SPIE) Conference Series, Vol. 12184,
  Ground-based and Airborne Instrumentation for Astronomy IX, ed. C.~J.
  {Evans}, J.~J. {Bryant}, \& K.~{Motohara}, 121845B,
  \dodoi{10.1117/12.2629090}

\bibitem[{{Lamarre} {et~al.}(2010){Lamarre}, {Puget}, {Ade}, {Bouchet},
  {Guyot}, {Lange}, {Pajot}, {Arondel}, {Benabed}, {Beney}, {Beno{\^\i}t},
  {Bernard}, {Bhatia}, {Blanc}, {Bock}, {Br{\'e}elle}, {Bradshaw}, {Camus},
  {Catalano}, {Charra}, {Charra}, {Church}, {Couchot}, {Coulais}, {Crill},
  {Crook}, {Dassas}, {de Bernardis}, {Delabrouille}, {de Marcillac}, {Delouis},
  {D{\'e}sert}, {Dumesnil}, {Dupac}, {Efstathiou}, {Eng}, {Evesque},
  {Fourmond}, {Ganga}, {Giard}, {Gispert}, {Guglielmi}, {Haissinski},
  {Henrot-Versill{\'e}}, {Hivon}, {Holmes}, {Jones}, {Koch}, {Lagard{\`e}re},
  {Lami}, {Land{\'e}}, {Leriche}, {Leroy}, {Longval}, {Mac{\'\i}as-P{\'e}rez},
  {Maciaszek}, {Maffei}, {Mansoux}, {Marty}, {Masi}, {Mercier},
  {Miville-Desch{\^e}nes}, {Moneti}, {Montier}, {Murphy}, {Narbonne}, {Nexon},
  {Paine}, {Pahn}, {Perdereau}, {Piacentini}, {Piat}, {Plaszczynski},
  {Pointecouteau}, {Pons}, {Ponthieu}, {Prunet}, {Rambaud}, {Recouvreur},
  {Renault}, {Ristorcelli}, {Rosset}, {Santos}, {Savini}, {Serra}, {Stassi},
  {Sudiwala}, {Sygnet}, {Tauber}, {Torre}, {Tristram}, {Vibert}, {Woodcraft},
  {Yurchenko}, \& {Yvon}}]{lamarre2010}
{Lamarre}, J.~M., {Puget}, J.~L., {Ade}, P.~A.~R., {et~al.} 2010, \aap, 520,
  A9, \dodoi{10.1051/0004-6361/200912975}

\bibitem[{{Larson}(1981)}]{larson1981}
{Larson}, R.~B. 1981, \mnras, 194, 809, \dodoi{10.1093/mnras/194.4.809}

\bibitem[{{Lee} {et~al.}(2004){Lee}, {Myers}, \& {Plume}}]{lee2004}
{Lee}, C.~W., {Myers}, P.~C., \& {Plume}, R. 2004, \apjs, 153, 523,
  \dodoi{10.1086/421996}

\bibitem[{{Leslie} {et~al.}(2020){Leslie}, {Schinnerer}, {Liu}, {Magnelli},
  {Algera}, {Karim}, {Davidzon}, {Gozaliasl}, {Jim{\'e}nez-Andrade}, {Lang},
  {Sargent}, {Novak}, {Groves}, {Smol{\v{c}}i{\'c}}, {Zamorani}, {Vaccari},
  {Battisti}, {Vardoulaki}, {Peng}, \& {Kartaltepe}}]{Leslie2020}
{Leslie}, S.~K., {Schinnerer}, E., {Liu}, D., {et~al.} 2020, \apj, 899, 58,
  \dodoi{10.3847/1538-4357/aba044}

\bibitem[{{Li} \& {Goldsmith}(2003)}]{Li2003}
{Li}, D., \& {Goldsmith}, P.~F. 2003, \apj, 585, 823, \dodoi{10.1086/346227}

\bibitem[{{Li} {et~al.}(2008){Li}, {Dowell}, {Kirby}, {Novak}, \&
  {Vaillancourt}}]{Li2008}
{Li}, H., {Dowell}, C.~D., {Kirby}, L., {Novak}, G., \& {Vaillancourt}, J.~E.
  2008, \ao, 47, 422, \dodoi{10.1364/AO.47.000422}

\bibitem[{{Li}(2017)}]{2017tcsf.book.....L}
{Li}, H.-b. 2017, {The Tai Chi in Star Formation},
  \dodoi{10.1088/978-1-6817-4293-9}

\bibitem[{{Mac Low} {et~al.}(2017){Mac Low}, {Burkert}, \&
  {Ib{\'a}{\~n}ez-Mej{\'\i}a}}]{MacLow2017}
{Mac Low}, M.-M., {Burkert}, A., \& {Ib{\'a}{\~n}ez-Mej{\'\i}a}, J.~C. 2017,
  \apjl, 847, L10, \dodoi{10.3847/2041-8213/aa8a61}

\bibitem[{{Mac Low} \& {Klessen}(2004)}]{maclow2004}
{Mac Low}, M.-M., \& {Klessen}, R.~S. 2004, Reviews of Modern Physics, 76, 125,
  \dodoi{10.1103/RevModPhys.76.125}

\bibitem[{{Magalhaes} {et~al.}(1996){Magalhaes}, {Rodrigues}, {Margoniner},
  {Pereyra}, \& {Heathcote}}]{magalhaes1996}
{Magalhaes}, A.~M., {Rodrigues}, C.~V., {Margoniner}, V.~E., {Pereyra}, A., \&
  {Heathcote}, S. 1996, in Astronomical Society of the Pacific Conference
  Series, Vol.~97, Polarimetry of the Interstellar Medium, ed. W.~G. {Roberge}
  \& D.~C.~B. {Whittet}, 118

\bibitem[{{Maharana} {et~al.}(2021){Maharana}, {Kypriotakis}, {Ramaprakash},
  {Rajarshi}, {Anche}, {Shrish}, {Blinov}, {Eriksen}, {Ghosh}, {Gjerl{\o}w},
  {Mandarakas}, {Panopoulou}, {Pavlidou}, {Pearson}, {Pelgrims}, {Potter},
  {Readhead}, {Skalidis}, {Tassis}, \& {Wehus}}]{Maharan2021}
{Maharana}, S., {Kypriotakis}, J.~A., {Ramaprakash}, A.~N., {et~al.} 2021,
  Journal of Astronomical Telescopes, Instruments, and Systems, 7, 014004,
  \dodoi{10.1117/1.JATIS.7.1.014004}

\bibitem[{{Maharana} {et~al.}(2022){Maharana}, {Anche}, {Ramaprakash}, {Joshi},
  {Basyrov}, {Blinov}, {Casadio}, {Deka}, {Eriksen}, {Ghosh}, {Gjerl{\o}w},
  {Kypriotakis}, {Kiehlmann}, {Mandarakas}, {Panopoulou}, {Papadaki},
  {Pavlidou}, {Pearson}, {Pelgrims}, {Potter}, {Readhead}, {Skalidis},
  {Svalheim}, {Tassis}, \& {Wehus}}]{Maharana2022}
{Maharana}, S., {Anche}, R.~M., {Ramaprakash}, A.~N., {et~al.} 2022, Journal of
  Astronomical Telescopes, Instruments, and Systems, 8, 038004,
  \dodoi{10.1117/1.JATIS.8.3.038004}

\bibitem[{{Mangilli} {et~al.}(2018){Mangilli}, {Fo{\"e}nard}, {Aumont},
  {Hughes}, {Mot}, {Bernard}, {Lacourt}, {Ristorcelli}, {Longval}, {Ade},
  {Andr{\'e}}, {Bautista}, {deBernardis}, {Boulade}, {Bousqet}, {Bouzit},
  {Buttice}, {Charra}, {Crane}, {Doumayrou}, {Dubois}, {Engel}, {Griffin},
  {Grabarnik}, {Hargrave}, {Laureijs}, {Leriche}, {Maestre}, {Maffei}, {Marty},
  {Marty}, {Masi}, {Misawa}, {Montel}, {Montier}, {Narbonne}, {Pajot},
  {P{\'e}rot}, {Pimentao}, {Pisano}, {Ponthieu}, {Rodriguez}, {Roudil},
  {Salatino}, {Savini}, {Simonella}, {Saccoccio}, {Stever}, {Tauber}, {Tibbs},
  \& {Tucker}}]{foenard2018}
{Mangilli}, A., {Fo{\"e}nard}, G., {Aumont}, J., {et~al.} 2018, arXiv e-prints,
  arXiv:1804.05645, \dodoi{10.48550/arXiv.1804.05645}

\bibitem[{{Marchal} \& {Miville-Desch{\^e}nes}(2021)}]{MarchalEA2021}
{Marchal}, A., \& {Miville-Desch{\^e}nes}, M.-A. 2021, \apj, 908, 186,
  \dodoi{10.3847/1538-4357/abd108}

\bibitem[{{Mathew} \& {Federrath}(2021)}]{MF2021}
{Mathew}, S.~S., \& {Federrath}, C. 2021, \mnras, 507, 2448,
  \dodoi{10.1093/mnras/stab2338}

\bibitem[{{Mathew} {et~al.}(2022){Mathew}, {Federrath}, \& {Seta}}]{MFS2022}
{Mathew}, S.~S., {Federrath}, C., \& {Seta}, A. 2022, \mnras,
  \dodoi{10.1093/mnras/stac3415}

\bibitem[{{Mestel}(1966)}]{mestel1966}
{Mestel}, L. 1966, \mnras, 133, 265, \dodoi{10.1093/mnras/133.2.265}

\bibitem[{{Murray} {et~al.}(1997){Murray}, {Nartallo}, {Haynes}, {Gannaway}, \&
  {Ade}}]{murray1997}
{Murray}, A.~G., {Nartallo}, R., {Haynes}, C.~V., {Gannaway}, F., \& {Ade},
  P.~A.~R. 1997, in ESA Special Publication, Vol. 401, The Far Infrared and
  Submillimetre Universe., ed. A.~{Wilson}, 405

\bibitem[{{Murray} {et~al.}(2015){Murray}, {Stanimirovi{\'c}}, {Goss},
  {Dickey}, {Heiles}, {Lindner}, {Babler}, {Pingel}, {Lawrence}, {Jencson}, \&
  {Hennebelle}}]{MurrayEA2015}
{Murray}, C.~E., {Stanimirovi{\'c}}, S., {Goss}, W.~M., {et~al.} 2015, \apj,
  804, 89, \dodoi{10.1088/0004-637X/804/2/89}

\bibitem[{{Nandakumar} \& {Dutta}(2023)}]{NandakumarD2023}
{Nandakumar}, M., \& {Dutta}, P. 2023, \mnras, 526, 4690,
  \dodoi{10.1093/mnras/stad3042}

\bibitem[{{Neha} {et~al.}(2018){Neha}, {Maheswar}, {Soam}, \& {Lee}}]{neha2018}
{Neha}, S., {Maheswar}, G., {Soam}, A., \& {Lee}, C.~W. 2018, \mnras, 476,
  4442, \dodoi{10.1093/mnras/sty485}

\bibitem[{{Neha} {et~al.}(2016){Neha}, {Maheswar}, {Soam}, {Lee}, \&
  {Tej}}]{neha2016}
{Neha}, S., {Maheswar}, G., {Soam}, A., {Lee}, C.~W., \& {Tej}, A. 2016, \aap,
  588, A45, \dodoi{10.1051/0004-6361/201526845}

\bibitem[{{Novak} {et~al.}(1989){Novak}, {Gonatas}, {Hildebrand}, \&
  {Platt}}]{novak1989}
{Novak}, G., {Gonatas}, D.~P., {Hildebrand}, R.~H., \& {Platt}, S.~R. 1989,
  \pasp, 101, 215, \dodoi{10.1086/132425}

\bibitem[{{Orkisz} {et~al.}(2017){Orkisz}, {Pety}, {Gerin}, {Bron},
  {Guzm{\'a}n}, {Bardeau}, {Goicoechea}, {Gratier}, {Le Petit}, {Levrier},
  {Liszt}, {{\"O}berg}, {Peretto}, {Roueff}, {Sievers}, \&
  {Tremblin}}]{Orkisz2017}
{Orkisz}, J.~H., {Pety}, J., {Gerin}, M., {et~al.} 2017, \aap, 599, A99,
  \dodoi{10.1051/0004-6361/201629220}

\bibitem[{{Otal}(2014)}]{otal2014}
{Otal}, L.~E. 2014, PhD thesis, Rheinische Friedrich Wilhelms University of
  Bonn, Germany

\bibitem[{{Pattle} {et~al.}(2022){Pattle}, {Fissel}, {Tahani}, {Liu}, \&
  {Ntormousi}}]{pattle2022}
{Pattle}, K., {Fissel}, L., {Tahani}, M., {Liu}, T., \& {Ntormousi}, E. 2022,
  arXiv e-prints, arXiv:2203.11179, \dodoi{10.48550/arXiv.2203.11179}

\bibitem[{{Pavel} \& {Clemens}(2012)}]{pavel2012}
{Pavel}, M.~D., \& {Clemens}, D.~P. 2012, \apj, 760, 150,
  \dodoi{10.1088/0004-637X/760/2/150}

\bibitem[{{Pernechele} {et~al.}(2012){Pernechele}, {Abe}, {Bendjoya},
  {Cellino}, {Massone}, {Rivet}, \& {Tanga}}]{Pernechele2012}
{Pernechele}, C., {Abe}, L., {Bendjoya}, P., {et~al.} 2012, in Society of
  Photo-Optical Instrumentation Engineers (SPIE) Conference Series, Vol. 8446,
  Ground-based and Airborne Instrumentation for Astronomy IV, ed. I.~S.
  {McLean}, S.~K. {Ramsay}, \& H.~{Takami}, 84462H, \dodoi{10.1117/12.925933}

\bibitem[{{Planck Collaboration} {et~al.}(2016){Planck Collaboration}, {Adam},
  {Ade}, {Aghanim}, {Arnaud}, {Ashdown}, {Aumont}, {Baccigalupi}, {Banday},
  {Barreiro}, {Bartolo}, {Battaner}, {Benabed}, {Beno{\^\i}t},
  {Benoit-L{\'e}vy}, {Bernard}, {Bersanelli}, {Bertincourt}, {Bielewicz},
  {Bock}, {Bonavera}, {Bond}, {Borrill}, {Bouchet}, {Boulanger}, {Bucher},
  {Burigana}, {Calabrese}, {Cardoso}, {Catalano}, {Challinor}, {Chamballu},
  {Chiang}, {Christensen}, {Clements}, {Colombi}, {Colombo}, {Combet},
  {Couchot}, {Coulais}, {Crill}, {Curto}, {Cuttaia}, {Danese}, {Davies},
  {Davis}, {de Bernardis}, {de Rosa}, {de Zotti}, {Delabrouille}, {Delouis},
  {D{\'e}sert}, {Diego}, {Dole}, {Donzelli}, {Dor{\'e}}, {Douspis}, {Ducout},
  {Dupac}, {Efstathiou}, {Elsner}, {En{\ss}lin}, {Eriksen}, {Falgarone},
  {Fergusson}, {Finelli}, {Forni}, {Frailis}, {Fraisse}, {Franceschi},
  {Frejsel}, {Galeotta}, {Galli}, {Ganga}, {Ghosh}, {Giard},
  {Giraud-H{\'e}raud}, {Gjerl{\o}w}, {Gonz{\'a}lez-Nuevo}, {G{\'o}rski},
  {Gratton}, {Gruppuso}, {Gudmundsson}, {Hansen}, {Hanson}, {Harrison},
  {Henrot-Versill{\'e}}, {Herranz}, {Hildebrandt}, {Hivon}, {Hobson}, {Holmes},
  {Hornstrup}, {Hovest}, {Huffenberger}, {Hurier}, {Jaffe}, {Jaffe}, {Jones},
  {Juvela}, {Keih{\"a}nen}, {Keskitalo}, {Kisner}, {Kneissl}, {Knoche}, {Kunz},
  {Kurki-Suonio}, {Lagache}, {Lamarre}, {Lasenby}, {Lattanzi}, {Lawrence}, {Le
  Jeune}, {Leahy}, {Lellouch}, {Leonardi}, {Lesgourgues}, {Levrier}, {Liguori},
  {Lilje}, {Linden-V{\o}rnle}, {L{\'o}pez-Caniego}, {Lubin},
  {Mac{\'\i}as-P{\'e}rez}, {Maggio}, {Maino}, {Mandolesi}, {Mangilli}, {Maris},
  {Martin}, {Mart{\'\i}nez-Gonz{\'a}lez}, {Masi}, {Matarrese}, {McGehee},
  {Melchiorri}, {Mendes}, {Mennella}, {Migliaccio}, {Mitra},
  {Miville-Desch{\^e}nes}, {Moneti}, {Montier}, {Moreno}, {Morgante},
  {Mortlock}, {Moss}, {Mottet}, {Munshi}, {Murphy}, {Naselsky}, {Nati},
  {Natoli}, {Netterfield}, {N{\o}rgaard-Nielsen}, {Noviello}, {Novikov},
  {Novikov}, {Oxborrow}, {Paci}, {Pagano}, {Pajot}, {Paoletti}, {Pasian},
  {Patanchon}, {Pearson}, {Perdereau}, {Perotto}, {Perrotta}, {Pettorino},
  {Piacentini}, {Piat}, {Pierpaoli}, {Pietrobon}, {Plaszczynski},
  {Pointecouteau}, {Polenta}, {Pratt}, {Pr{\'e}zeau}, {Prunet}, {Puget},
  {Rachen}, {Reinecke}, {Remazeilles}, {Renault}, {Renzi}, {Ristorcelli},
  {Rocha}, {Rosset}, {Rossetti}, {Roudier}, {Rusholme}, {Sandri}, {Santos},
  {Sauv{\'e}}, {Savelainen}, {Savini}, {Scott}, {Seiffert}, {Shellard},
  {Spencer}, {Stolyarov}, {Stompor}, {Sudiwala}, {Sutton}, {Suur-Uski},
  {Sygnet}, {Tauber}, {Terenzi}, {Toffolatti}, {Tomasi}, {Tristram}, {Tucci},
  {Tuovinen}, {Valenziano}, {Valiviita}, {Van Tent}, {Vibert}, {Vielva},
  {Villa}, {Wade}, {Wandelt}, {Watson}, {Wehus}, {Yvon}, {Zacchei}, \&
  {Zonca}}]{planckColVIII2016}
{Planck Collaboration}, {Adam}, R., {Ade}, P.~A.~R., {et~al.} 2016, \aap, 594,
  A8, \dodoi{10.1051/0004-6361/201525820}

\bibitem[{{Platt}(1991)}]{Platt1991}
{Platt}, S.~R. 1991, PhD thesis, University of Chicago

\bibitem[{{Primiani} {et~al.}(2016){Primiani}, {Young}, {Young}, {Patel},
  {Wilson}, {Vertatschitsch}, {Chitwood}, {Srinivasan}, {MacMahon}, \&
  {Weintroub}}]{primiani2016}
{Primiani}, R.~A., {Young}, K.~H., {Young}, A., {et~al.} 2016, Journal of
  Astronomical Instrumentation, 5, 1641006, \dodoi{10.1142/S2251171716410063}

\bibitem[{{Rai} {et~al.}(2020){Rai}, {Ganesh}, {Paul}, {Kasarla}, {Prajapati},
  {Sarkar}, {Singh}, {Patwal}, {Adalja}, {Mathur}, {Naik}, {Shah}, \&
  {Baliyan}}]{Archita2020}
{Rai}, A., {Ganesh}, S., {Paul}, S.~K., {et~al.} 2020, in Society of
  Photo-Optical Instrumentation Engineers (SPIE) Conference Series, Vol. 11447,
  Ground-based and Airborne Instrumentation for Astronomy VIII, ed. C.~J.
  {Evans}, J.~J. {Bryant}, \& K.~{Motohara}, 1144765,
  \dodoi{10.1117/12.2560988}

\bibitem[{{Ramaprakash} {et~al.}(1998){Ramaprakash}, {Gupta}, {Sen}, \&
  {Tandon}}]{ram1998}
{Ramaprakash}, A.~N., {Gupta}, R., {Sen}, A.~K., \& {Tandon}, S.~N. 1998,
  \aaps, 128, 369, \dodoi{10.1051/aas:1998148}

\bibitem[{{Rautela} {et~al.}(2004){Rautela}, {Joshi}, \&
  {Pandey}}]{rautela2004}
{Rautela}, B.~S., {Joshi}, G.~C., \& {Pandey}, J.~C. 2004, Bulletin of the
  Astronomical Society of India, 32, 159

\bibitem[{{Renbarger} {et~al.}(2004){Renbarger}, {Chuss}, {Dotson}, {Griffin},
  {Hanna}, {Loewenstein}, {Malhotra}, {Marshall}, {Novak}, \&
  {Pernic}}]{renbarger2004}
{Renbarger}, T., {Chuss}, D.~T., {Dotson}, J.~L., {et~al.} 2004, \pasp, 116,
  415, \dodoi{10.1086/383623}

\bibitem[{{Ritacco} {et~al.}(2020){Ritacco}, {Adam}, {Ade}, {Ajeddig},
  {Andr{\'e}}, {Andrianasolo}, {Aussel}, {Beelen}, {Beno{\^\i}t}, {Bideaud},
  {Bourrion}, {Calvo}, {Catalano}, {Comis}, {De Petris}, {D{\'e}sert}, {Doyle},
  {Driessen}, {Gomez}, {Goupy}, {K{\'e}ruzor{\'e}}, {Kramer}, {Ladjelate},
  {Lagache}, {Leclercq}, {Lestrade}, {Mac{\'\i}as-P{\'e}rez}, {Mauskopf},
  {Maury}, {Mayet}, {Monfardini}, {Perotto}, {Pisano}, {Ponthieu},
  {Rev{\'e}ret}, {Romero}, {Roussel}, {Ruppin}, {Schuster}, {Shimajiri}, {Shu},
  {Sievers}, {Tucker}, \& {Zylka}}]{ritacco2020}
{Ritacco}, A., {Adam}, R., {Ade}, P., {et~al.} 2020, in European Physical
  Journal Web of Conferences, Vol. 228, European Physical Journal Web of
  Conferences, 00022, \dodoi{10.1051/epjconf/202022800022}

\bibitem[{{Roelfsema} {et~al.}(2018){Roelfsema}, {Shibai}, {Armus}, {Arrazola},
  {Audard}, {Audley}, {Bradford}, {Charles}, {Dieleman}, {Doi}, {Duband},
  {Eggens}, {Evers}, {Funaki}, {Gao}, {Giard}, {di Giorgio}, {Gonz{\'a}lez
  Fern{\'a}ndez}, {Griffin}, {Helmich}, {Hijmering}, {Huisman}, {Ishihara},
  {Isobe}, {Jackson}, {Jacobs}, {Jellema}, {Kamp}, {Kaneda}, {Kawada},
  {Kemper}, {Kerschbaum}, {Khosropanah}, {Kohno}, {Kooijman}, {Krause}, {van
  der Kuur}, {Kwon}, {Laauwen}, {de Lange}, {Larsson}, {van Loon}, {Madden},
  {Matsuhara}, {Najarro}, {Nakagawa}, {Naylor}, {Ogawa}, {Onaka}, {Oyabu},
  {Poglitsch}, {Reveret}, {Rodriguez}, {Spinoglio}, {Sakon}, {Sato},
  {Shinozaki}, {Shipman}, {Sugita}, {Suzuki}, {van der Tak}, {Torres Redondo},
  {Wada}, {Wang}, {Wafelbakker}, {van Weers}, {Withington}, {Vandenbussche},
  {Yamada}, \& {Yamamura}}]{roelfsema2018}
{Roelfsema}, P.~R., {Shibai}, H., {Armus}, L., {et~al.} 2018, \pasa, 35, e030,
  \dodoi{10.1017/pasa.2018.15}

\bibitem[{{Saha} {et~al.}(2021){Saha}, {Gopinathan}, {Sharma}, {Won Lee},
  {Ghosh}, \& {Kim}}]{saha2021}
{Saha}, P., {Gopinathan}, M., {Sharma}, E., {et~al.} 2021, \aap, 655, A76,
  \dodoi{10.1051/0004-6361/202039948}

\bibitem[{{Saha} {et~al.}(2022){Saha}, {Soam}, {Baug}, {Gopinathan}, {Mondal},
  \& {Ghosh}}]{saha2022}
{Saha}, P., {Soam}, A., {Baug}, T., {et~al.} 2022, \mnras, 513, 2039,
  \dodoi{10.1093/mnras/stac943}

\bibitem[{{Schleuning} {et~al.}(1997){Schleuning}, {Dowell}, {Hildebrand},
  {Platt}, \& {Novak}}]{schleuning1997}
{Schleuning}, D.~A., {Dowell}, C.~D., {Hildebrand}, R.~H., {Platt}, S.~R., \&
  {Novak}, G. 1997, \pasp, 109, 307, \dodoi{10.1086/133892}

\bibitem[{{Scowen} {et~al.}(2022){Scowen}, {Jones}, \& {Oudmaijer}}]{Paul2022}
{Scowen}, P.~A., {Jones}, C.~E., \& {Oudmaijer}, R.~D. 2022, \apss, 367, 126,
  \dodoi{10.1007/s10509-022-04143-5}

\bibitem[{{Seta} \& {Federrath}(2022)}]{SF2022}
{Seta}, A., \& {Federrath}, C. 2022, \mnras, 514, 957,
  \dodoi{10.1093/mnras/stac1400}

\bibitem[{{Sharma} {et~al.}(2022){Sharma}, {Gopinathan}, {Soam}, {Lee}, \&
  {Seshadri}}]{ekta2022}
{Sharma}, E., {Gopinathan}, M., {Soam}, A., {Lee}, C.~W., \& {Seshadri}, T.~R.
  2022, \mnras, 517, 1138, \dodoi{10.1093/mnras/stac2487}

\bibitem[{{Silpa} {et~al.}(2021){Silpa}, {Kharb}, {Harrison}, {Ho}, {Jarvis},
  {Ishwara-Chandra}, \& {Sebastian}}]{Shilpa2021}
{Silpa}, S., {Kharb}, P., {Harrison}, C.~M., {et~al.} 2021, \mnras, 507, 991,
  \dodoi{10.1093/mnras/stab1870}

\bibitem[{{Siringo} {et~al.}(2012){Siringo}, {Kov{\'a}cs}, {Kreysa},
  {Schuller}, {Weiss}, {Guesten}, {Hezareh}, {Menten}, {Wiesemeyer}, {Dumke},
  {Montenegro}, \& {Parra}}]{siringo2012}
{Siringo}, G., {Kov{\'a}cs}, A., {Kreysa}, E., {et~al.} 2012, in Society of
  Photo-Optical Instrumentation Engineers (SPIE) Conference Series, Vol. 8452,
  Millimeter, Submillimeter, and Far-Infrared Detectors and Instrumentation for
  Astronomy VI, ed. W.~S. {Holland} \& J.~{Zmuidzinas}, 845206,
  \dodoi{10.1117/12.925697}

\bibitem[{{Soam} {et~al.}(2015){Soam}, {Maheswar}, {Lee}, {Dib}, {Bhatt},
  {Tamura}, \& {Kim}}]{soam2015}
{Soam}, A., {Maheswar}, G., {Lee}, C.~W., {et~al.} 2015, \aap, 573, A34,
  \dodoi{10.1051/0004-6361/201322536}

\bibitem[{{Soam} {et~al.}(2019){Soam}, {Lee}, {Andersson}, {Maheswar},
  {Juvela}, {Liu}, {Kim}, {Rao}, {Chung}, {Kwon}, \& {Ekta}}]{soam2019}
{Soam}, A., {Lee}, C.~W., {Andersson}, B.~G., {et~al.} 2019, \apj, 883, 9,
  \dodoi{10.3847/1538-4357/ab365d}

\bibitem[{{Soam} {et~al.}(2021){Soam}, {Andersson}, {Strai{\v{z}}ys}, {Caputo},
  {Kazlauskas}, {Boyle}, {Janusz}, {Zdanavi{\v{c}}ius}, \&
  {Acosta-Pulido}}]{soam2021}
{Soam}, A., {Andersson}, B.~G., {Strai{\v{z}}ys}, V., {et~al.} 2021, \aj, 161,
  149, \dodoi{10.3847/1538-3881/abdd3b}

\bibitem[{{Staguhn} {et~al.}(2018){Staguhn}, {Amatucci}, {Armus}, {Bradley},
  {Carter}, {Chuss}, {Corsetti}, {Cooray}, {Howard}, {Leisawitz}, {Meixner},
  {Moseley}, {Pope}, {Vieira}, \& {Wollack}}]{staguhn2018}
{Staguhn}, J., {Amatucci}, E., {Armus}, L., {et~al.} 2018, in Society of
  Photo-Optical Instrumentation Engineers (SPIE) Conference Series, Vol. 10698,
  Space Telescopes and Instrumentation 2018: Optical, Infrared, and Millimeter
  Wave, ed. M.~{Lystrup}, H.~A. {MacEwen}, G.~G. {Fazio}, N.~{Batalha},
  N.~{Siegler}, \& E.~C. {Tong}, 106981A, \dodoi{10.1117/12.2312626}

\bibitem[{{Sutin} {et~al.}(2018){Sutin}, {Alvarez}, {Battaglia}, {Bock},
  {Bonato}, {Borrill}, {Chuss}, {Cooperrider}, {Crill}, {Delabrouille},
  {Devlin}, {Essinger-Hileman}, {Fissel}, {Flauger}, {Gorski}, {Green},
  {Hanany}, {Hubmayr}, {Johnson}, {Jones}, {Knox}, {Kogut}, {Lawrence},
  {McMahon}, {Matsumura}, {Negrello}, {O'Brient}, {Paine}, {Pryke}, {Shirron},
  {Trangsrud}, {Wen}, {Young}, \& {de Zotti}}]{sutin2018}
{Sutin}, B.~M., {Alvarez}, M., {Battaglia}, N., {et~al.} 2018, in Society of
  Photo-Optical Instrumentation Engineers (SPIE) Conference Series, Vol. 10698,
  Space Telescopes and Instrumentation 2018: Optical, Infrared, and Millimeter
  Wave, ed. M.~{Lystrup}, H.~A. {MacEwen}, G.~G. {Fazio}, N.~{Batalha},
  N.~{Siegler}, \& E.~C. {Tong}, 106984F, \dodoi{10.1117/12.2311326}

\bibitem[{{Tafalla} {et~al.}(1998){Tafalla}, {Mardones}, {Myers}, {Caselli},
  {Bachiller}, \& {Benson}}]{Tafalla1998}
{Tafalla}, M., {Mardones}, D., {Myers}, P.~C., {et~al.} 1998, \apj, 504, 900,
  \dodoi{10.1086/306115}

\bibitem[{{Tahani} {et~al.}(2023){Tahani}, {Bastien}, {Furuya}, {Pattle},
  {Johnstone}, {Arzoumanian}, {Doi}, {Hasegawa}, {Inutsuka}, {Coud{\'e}},
  {Fissel}, {Chen}, {Poidevin}, {Sadavoy}, {Friesen}, {Koch}, {Di Francesco},
  {Moriarty-Schieven}, {Chen}, {Chung}, {Eswaraiah}, {Fanciullo}, {Gledhill},
  {Le Gouellec}, {Hoang}, {Hwang}, {Kang}, {Kim}, {Kirchschlager}, {Kwon},
  {Lee}, {Liu}, {Onaka}, {Rawlings}, {Soam}, {Tamura}, {Tang}, {Tomisaka},
  {Whitworth}, {Kwon}, {Hoang}, {Redman}, {Berry}, {Ching}, {Wang}, {Lai},
  {Qiu}, {Ward-Thompson}, {Houde}, {Byun}, {Chen}, {Chen}, {Cho}, {Choi},
  {Choi}, {Chrysostomou}, {Diep}, {Duan}, {Fiege}, {Franzmann}, {Friberg},
  {Fuller}, {Graves}, {Greaves}, {Griffin}, {Gu}, {Han}, {Hatchell}, {Hayashi},
  {Hull}, {Inoue}, {Iwasaki}, {Jeong}, {Kanamori}, {Kang}, {Kang}, {Kataoka},
  {Kawabata}, {Kemper}, {Kim}, {Kim}, {Kim}, {Kim}, {Kim}, {Kirk}, {Kobayashi},
  {Konyves}, {Kusune}, {Lacaille}, {Law}, {Lee}, {Lee}, {Lee}, {Lee}, {Lee},
  {Li}, {Li}, {Li}, {Liu}, {Liu}, {Liu}, {de Looze}, {Lyo}, {Mairs},
  {Matsumura}, {Matthews}, {Nagata}, {Nakamura}, {Nakanishi}, {Ohashi}, {Park},
  {Parsons}, {Peretto}, {Pyo}, {Qian}, {Rao}, {Retter}, {Richer}, {Rigby},
  {Saito}, {Savini}, {Scaife}, {Seta}, {Shimajiri}, {Shinnaga}, {Tang},
  {Tsukamoto}, {Viti}, {Wang}, {Yen}, {Yoo}, {Yuan}, {Yun}, {Zenko}, {Zhang},
  {Zhang}, {Zhang}, {Zhou}, {Zhu}, {Andr{\'e}}, {Dowell}, {Eyres}, {Falle},
  {van Loo}, \& {Robitaille}}]{tahani2023}
{Tahani}, M., {Bastien}, P., {Furuya}, R.~S., {et~al.} 2023, \apj, 944, 139,
  \dodoi{10.3847/1538-4357/acac81}

\bibitem[{{Thum} {et~al.}(2008){Thum}, {Wiesemeyer}, {Paubert}, {Navarro}, \&
  {Morris}}]{thum2008}
{Thum}, C., {Wiesemeyer}, H., {Paubert}, G., {Navarro}, S., \& {Morris}, D.
  2008, \pasp, 120, 777, \dodoi{10.1086/590190}

\bibitem[{{V{\'a}zquez-Semadeni} {et~al.}(2019){V{\'a}zquez-Semadeni}, {Palau},
  {Ballesteros-Paredes}, {G{\'o}mez}, \& {Zamora-Avil{\'e}s}}]{semadeni2019}
{V{\'a}zquez-Semadeni}, E., {Palau}, A., {Ballesteros-Paredes}, J.,
  {G{\'o}mez}, G.~C., \& {Zamora-Avil{\'e}s}, M. 2019, \mnras, 490, 3061,
  \dodoi{10.1093/mnras/stz2736}

\bibitem[{{Ward-Thompson} {et~al.}(2017){Ward-Thompson}, {Pattle}, {Bastien},
  {Furuya}, {Kwon}, {Lai}, {Qiu}, {Berry}, {Choi}, {Coud{\'e}}, {Di Francesco},
  {Hoang}, {Franzmann}, {Friberg}, {Graves}, {Greaves}, {Houde}, {Johnstone},
  {Kirk}, {Koch}, {Kwon}, {Lee}, {Li}, {Matthews}, {Mottram}, {Parsons}, {Pon},
  {Rao}, {Rawlings}, {Shinnaga}, {Sadavoy}, {van Loo}, {Aso}, {Byun},
  {Eswaraiah}, {Chen}, {Chen}, {Chen}, {Ching}, {Cho}, {Chrysostomou}, {Chung},
  {Doi}, {Drabek-Maunder}, {Eyres}, {Fiege}, {Friesen}, {Fuller}, {Gledhill},
  {Griffin}, {Gu}, {Hasegawa}, {Hatchell}, {Hayashi}, {Holland}, {Inoue},
  {Inutsuka}, {Iwasaki}, {Jeong}, {Kang}, {Kang}, {Kang}, {Kawabata}, {Kemper},
  {Kim}, {Kim}, {Kim}, {Kim}, {Kim}, {Kim}, {Lacaille}, {Lee}, {Lee}, {Li},
  {Li}, {Liu}, {Liu}, {Liu}, {Liu}, {Lyo}, {Mairs}, {Matsumura},
  {Moriarty-Schieven}, {Nakamura}, {Nakanishi}, {Ohashi}, {Onaka}, {Peretto},
  {Pyo}, {Qian}, {Retter}, {Richer}, {Rigby}, {Robitaille}, {Savini}, {Scaife},
  {Soam}, {Tamura}, {Tang}, {Tomisaka}, {Wang}, {Wang}, {Whitworth}, {Yen},
  {Yoo}, {Yuan}, {Zhang}, {Zhang}, {Zhou}, {Zhu}, {Andr{\'e}}, {Dowell},
  {Falle}, \& {Tsukamoto}}]{ward-thompson2017}
{Ward-Thompson}, D., {Pattle}, K., {Bastien}, P., {et~al.} 2017, \apj, 842, 66,
  \dodoi{10.3847/1538-4357/aa70a0}

\bibitem[{{Wiesemeyer} {et~al.}(2014){Wiesemeyer}, {Hezareh}, {Kreysa},
  {Weiss}, {G{\"u}sten}, {Menten}, {Siringo}, {Schuller}, \&
  {Kovacs}}]{wiesemeyer2014}
{Wiesemeyer}, H., {Hezareh}, T., {Kreysa}, E., {et~al.} 2014, \pasp, 126, 1027,
  \dodoi{10.1086/679002}

\bibitem[{{Wilson} {et~al.}(2020){Wilson}, {Abi-Saad}, {Ade}, {Aretxaga},
  {Austermann}, {Ban}, {Bardin}, {Beall}, {Berthoud}, {Bryan}, {Bussan},
  {Castillo}, {Chavez}, {Contente}, {DeNigris}, {Dober}, {Eiben}, {Ferrusca},
  {Fissel}, {Gao}, {Golec}, {Golina}, {Gomez}, {Gordon}, {Gutermuth}, {Hilton},
  {Hosseini}, {Hubmayr}, {Hughes}, {Kuczarski}, {Lee}, {Lunde}, {Ma}, {Mani},
  {Mauskopf}, {McCrackan}, {McKenney}, {McMahon}, {Novak}, {Pisano}, {Pope},
  {Ralston}, {Rodriguez}, {S{\'a}nchez-Arg{\"u}elles}, {Schloerb}, {Simon},
  {Sinclair}, {Souccar}, {Torres Campos}, {Tucker}, {Ullom}, {Van Camp}, {Van
  Lanen}, {Velazquez}, {Vissers}, {Weeks}, \& {Yun}}]{wilson2020}
{Wilson}, G.~W., {Abi-Saad}, S., {Ade}, P., {et~al.} 2020, in Society of
  Photo-Optical Instrumentation Engineers (SPIE) Conference Series, Vol. 11453,
  Society of Photo-Optical Instrumentation Engineers (SPIE) Conference Series,
  1145302, \dodoi{10.1117/12.2562331}

\bibitem[{{Young} {et~al.}(2018){Young}, {Alvarez}, {Battaglia}, {Bock},
  {Borrill}, {Chuss}, {Crill}, {Delabrouille}, {Devlin}, {Fissel}, {Flauger},
  {Green}, {Gorski}, {Hanany}, {Hills}, {Hubmayr}, {Johnson}, {Jones}, {Knox},
  {Kogut}, {Lawrence}, {Matsumura}, {McGuire}, {McMahon}, {O'Brient}, {Pryke},
  {Sutin}, {Tan}, {Trangsrud}, {Wen}, \& {De Zotti}}]{young2018}
{Young}, K., {Alvarez}, M., {Battaglia}, N., {et~al.} 2018, in Society of
  Photo-Optical Instrumentation Engineers (SPIE) Conference Series, Vol. 10698,
  Space Telescopes and Instrumentation 2018: Optical, Infrared, and Millimeter
  Wave, ed. M.~{Lystrup}, H.~A. {MacEwen}, G.~G. {Fazio}, N.~{Batalha},
  N.~{Siegler}, \& E.~C. {Tong}, 1069846, \dodoi{10.1117/12.2309421}

\bibitem[{{Zhang} {et~al.}(2014){Zhang}, {Qiu}, {Girart}, {Liu}, {Tang},
  {Koch}, {Li}, {Keto}, {Ho}, {Rao}, {Lai}, {Ching}, {Frau}, {Chen}, {Li},
  {Padovani}, {Bontemps}, {Csengeri}, \& {Ju{\'a}rez}}]{Zhang2014}
{Zhang}, Q., {Qiu}, K., {Girart}, J.~M., {et~al.} 2014, \apj, 792, 116,
  \dodoi{10.1088/0004-637X/792/2/116}

\end{thebibliography}
\end{footnotesize}

\end{document}